\documentclass[11pt]{article}
\usepackage{changepage}
\usepackage[utf8]{inputenc}
\usepackage{amsmath,amssymb}
\usepackage{cite}
\usepackage{appendix}
\usepackage{nameref,hyperref}

\usepackage[aboveskip=1pt,labelfont=bf,labelsep=period,justification=raggedright,singlelinecheck=off]{caption}

\date{}

\usepackage{graphicx}
\usepackage{dcolumn}
\usepackage{bm}
\usepackage{latexsym}
\usepackage{mathtools, gensymb, hyperref, dsfont, amssymb}
\usepackage{enumerate}
\usepackage{cancel}
\usepackage[caption=false]{subfig}

\usepackage{lastpage,graphicx}
\usepackage{epstopdf}

\newcommand{\fab}{f_{\text{AB}}}
\newcommand{\beeqn}{\begin{eqnarray*}}
\newcommand{\eeqn}{\end{eqnarray*}}

\newcommand{\ba}[1]{ \begin{align}#1\end{align} }
\newcommand{\bad}[1]{ \begin{aligned}#1\end{aligned} }
\newcommand{\beeqnn}{\begin{eqnarray}}
\newcommand{\eeqnn}{\end{eqnarray}}

\newcommand{\deriv}[2]{\frac{d#1}{d#2}}
\newcommand{\benum}{\begin{enumerate}[a)]}
\newcommand{\eenum}{\end{enumerate}}

\newcommand{\nn}{\nonumber}
\newcommand{\be}{\begin{equation}}
\newcommand{\ee}{\end{equation}}

\begin{document}

\begin{flushleft}
{\Large
\textbf\newline{Optimizing Real-Time Vaccine Allocation in a Stochastic SIR Model}
}
\newline
\\
Chantal Nguyen\textsuperscript{*},
Jean M. Carlson
\\
\bigskip
\bf Department of Physics, University of California Santa Barbara, Santa Barbara, California, United States of America
\\
\bigskip

* cnguyen@physics.ucsb.edu

\end{flushleft}
\section*{Abstract}
Real-time vaccination following an outbreak can effectively mitigate the damage caused by an infectious disease. However, in many cases, available resources are insufficient to vaccinate the entire at-risk population, logistics result in delayed vaccine deployment, and the interaction between members of different cities facilitates a wide spatial spread of infection. Limited vaccine, time delays, and interaction (or coupling) of cities lead to tradeoffs that impact the overall magnitude of the epidemic. These tradeoffs mandate investigation of optimal strategies that minimize the severity of the epidemic by prioritizing allocation of vaccine to specific subpopulations. We use an SIR model to describe the disease dynamics of an epidemic which breaks out in one city and spreads to another. We solve a master equation to determine the resulting probability distribution of the final epidemic size. We then identify tradeoffs between vaccine, time delay, and coupling, and we determine the optimal vaccination protocols resulting from these tradeoffs. 

\section*{Introduction}
Upon the outbreak of infectious disease, effective and widespread intervention through vaccination is of immediate concern. However, distributing and deploying vaccine relies on numerous logistical or even political factors that can result in delays. Meanwhile, the extensive modern network of rapid transportation facilitates the spread of infectious disease between communities. If the resources available are insufficient to fully immunize the entire population, strategically allocating vaccine to specific subpopulations can minimize the spread and severity of infection. The interaction between members of different cities and the time delay until vaccination lead to tradeoffs that can dictate optimal strategies of distributing limited vaccine. Optimizing resource allocation is of great interest to policymakers who must decide who gets vaccinated and when.

Real-time vaccination after an outbreak has occurred necessarily involves delays resulting from the production, testing, and/or delivery of vaccine. Such time delays can change the vaccine allocation strategy which results in the fewest infections, in addition to greatly impacting the severity of the epidemic.
During the 2009 H1N1 influenza outbreak, a widespread vaccination campaign in the United States successfully prevented between 0.7 and 1.5 million cases. However, had the campaign started one week earlier, it is estimated that approximately 27\% more cases could have been prevented, and had the campaign begun two weeks earlier, approximately 59\% more cases could have been prevented \cite{cdc}.

In some cases, stockpiles of vaccine have been established to expedite emergency intervention, but there still remain time delays between the initial outbreak and widespread control. For example, the International Coordinating Group (ICG) on Vaccine Provision for Epidemic Meningitis Control, a collaboration between the World Health Organization, UNICEF, M\'{e}decins Sans Fronti\`{e}res, and the International Federation of the Red Cross, stockpiles meningococcal vaccines as an emergency control method \cite{yen}. However, before deployment of stockpiled vaccines, the government or organization in need must first submit a request to the ICG, which then approves or disapproves the request within 48 hours; procurement and delivery of vaccine takes up to another 7 days. Furthermore, a multinational cooperative effort to undertake precautionary measures is not readily established \cite{fidler}. Developed nations usually have the means to produce and stockpile large quantities of vaccines for themselves in preparation for an epidemic, but developing nations often rely on vaccines donated by developed nations or acquired in exchange for virus samples. 
In the case of the 2009 H1N1 epidemic, developed and developing nations did not successfully broker a vaccine-sharing deal in time.
Moreover, some nations which have established stockpiles may keep vaccines for themselves during an epidemic. Fearing shortages at home, some nations may delay exporting their stockpiled vaccines (or refuse to export outright), another source of the time delay that hinders efficient, effective epidemic control \cite{fidler}.

In this paper, we use an SIR model to simulate the spread of infection and the effects of real-time vaccination. Commonly used in epidemiology, the SIR model divides a population into three compartments (susceptible $S$, infected $I$, and recovered $R$) with absorbing transitions between states $S$ and $I$ and $I$ and $R$ \cite{kermack}. 
Most SIR models are formulated as deterministic mean-field theories consisting of coupled differential equations, especially when vaccination is taken into account, since this can significantly complicate the model and increase computational intensity. Stochastic models, on the other hand, take into consideration the variability of the final epidemic size as well as the nonzero probability that infection fails to spread even in interacting communities. 

To investigate the disease dynamics, we utilize a discrete stochastic model complementary to the deterministic approach. Our stochastic model serves as a critical check on the conclusions drawn from the deterministic model. If an outbreak occurs within the deterministic model, without intervention, the infection will eventually spread to every population which interacts with the source of the infection. On the other hand, stochastic effects increase the temporal uncertainty in infectious spread between cities, and the disease may completely fail to spread from one city to another.

Rather than proactive (or prophylactic) vaccination, we focus on \textit{reactive} vaccination administered after the initial outbreak. The real-time response interacts with the dynamics of the epidemic, ultimately impacting the final outcome. The disease dynamics in interacting populations are not necessarily synchronous; infection can be spreading throughout one city while it is dying out in another. Simulating reactive vaccination is more computationally intensive than prophylaxis, necessitating limited population sizes in our simulations, since the state of the system must be determined at intermediate points during the epidemic, rather than simply the final end state.

In this paper, we first investigate the spread of disease within an individual population and identify tradeoffs between the amount of available vaccine and the time delay until vaccine is administered. We then examine the spread of infection between two interacting cities linked via implied transportation routes, and we investigate how coupling, vaccine, and time delay contribute to tradeoffs which in turn determine optimal vaccine allocation strategies. Finally, we compare the optimal protocols with those of the deterministic case, and we determine the differences between the optimal and worst-case scenarios.

\section*{Methods}
\subsection*{SIR Model}
In order to explore the temporal dynamics of an epidemic and investigate the effects of intervention methods, we utilize the mathematical framework of the SIR model \cite{kermack}, which has been commonly applied to infectious diseases such as influenza \cite{flu}, measles \cite{measles}, and whooping cough \cite{whoop}. In its simplest form, the SIR model divides a population of $N$ individuals into three compartments, $S$, $I$, and $R$, representing susceptible, infected, and recovered, respectively. A susceptible individual becomes infected only through contact with an infective, and an infective, once recovered, develops an immunity to the disease, thus permanently remaining in the recovered state. 

The processes of infection and recovery can be described as an analog to two chemical reactions, denoted $Z_1$ and $Z_2$, with reaction rates $\beta$ and $\gamma$, respectively:
\be \label{eq:reac} \bad{&Z_1: S + I \xrightarrow{\beta} 2I, \\ &Z_2: I \xrightarrow{\gamma} R.} \ee

Assuming a well-mixed population (i.e. an individual has equal probability of interaction with every other individual), coupled, ordinary differential equations can be used to describe the epidemic \cite{edwin}:
\begin{align} 
\deriv{S}{t} &= -\beta S I, \\ 
\deriv{I}{t} &= \beta S I - \gamma I, \\ 
\deriv{R}{t} &= \gamma I. \end{align} 
Susceptible individuals become infected at rate $\beta$, and infected individuals recover at rate $\gamma$. The last equation can be omitted if $N$ is kept fixed, and $R$ can be deduced from $R(t) = 1 - S(t) - I(t)$.

The infection rate $\beta$ is defined as the product of the average number of contacts each individual makes per unit time, $c$, and the probability of infection via contact, $p$, divided by the total population size, $N$ \cite{edwin}:
\be \beta = \frac{\text{contact rate} \times \text{probability of infection}}{\text{total population size}} = \frac{cp}{N}. \ee
The recovery rate $\gamma$ is simply the inverse of the characteristic timescale over which an individual remains infected $T$:
\be \gamma = \frac{1}{\text{average infection duration}} = \frac{1}{T}. \ee

We also define the reproductive number $r_0$, which represents the average number of individuals that will be infected by a single infective in a population completely consisting of susceptibles \cite{edwin}:
\be r_0 \equiv \frac{\beta S_0}{\gamma} = \frac{c p T S_0}{N}. \ee
where $S_0 = S(t=0)$, the initial fraction of susceptibles. If $r_0 < 1$, then in the deterministic model, $\frac{dI}{dt} < 0$; hence, the number of infectives declines from the individual value and no epidemic will occur. The value of $r_0$ varies greatly depending on the disease; $r_0$ is approximately 1.7 for influenza \cite{flu} and approximately 17 for measles and whooping cough \cite{kr}.

\subsection*{Stochasticity and the Master Equation}
Epidemics are not continuous processes, since $S$, $I$, and $R$ levels can only change by discrete integer values. Deterministic models are typically continuous mean-field theories that, unlike stochastic models, do not account for variability in epidemic sizes. When populations are weakly coupled, stochasticity can result in epidemics spreading from host cities to uninfected cities over a longer time frame than in purely deterministic models \cite{kr}. Furthermore, stochastic models include the nonzero probability that the epidemic does not progress. The probability distribution of the cumulative number infected at asymptotically infinite time is typically bimodal, with one peak representing a large-scale epidemic, and one peak representing the failure of the infection to significantly spread. Moreover, in a discrete stochastic model, positive integer numbers of individuals must make contact in order to transmit disease, while a continuous deterministic model does not include this requirement.

In this paper, we model epidemics as semi-Markovian processes and investigate the role of stochasticity in epidemics in coupled cities. The time evolution of the probability distribution is described by a master equation. Defining $P_{(S,I)} = P_{(S,I)}(t)$ as the probability of being in state ($S, I$) at time $t$, the SIR master equation takes the form \cite{kr}:
\be \begin{split} \label{eq:master} \frac{d P_{(S,I)}}{dt} = &\ \beta (S+1)(I-1) P_{(S+1,I-1)} \\ &+ \gamma(I+1)P_{(S,I+1)} - (\beta SI + \gamma I)P_{(S,I)}.\end{split} \ee 
The first term on the RHS represents a transition into state ($S,I$) by a susceptible individual becoming infected; the second term represents a transition into state ($S,I$) by an infected individual recovering; and the third term represents leaving the state ($S, I$) by infection or recovery. Only one reaction can occur in each infinitesimal time interval $[t, t+dt]$. 

The master equation is linear and can be written in matrix form with a column vector $\vec{P}$ containing probabilities of all possible states and a transition rate (or generator) matrix $\mathds{A}$ containing the coefficients of Eq.\@ \ref{eq:master}:
\be \frac{d\vec{P}}{dt} = \mathds{A} \vec{P}. \ee

Rather than calculating $\vec{P}$ by counting $S$ and $I$ levels, we follow the ``degree of advancement'' (DA) process \cite{jg} and count the occurrences of the $Z_1$ and $Z_2$ reactions described in Eq. \ref{eq:reac}. The vector $\vec{P}(t)$ is composed of probabilities of each possible state $(Z_1,Z_2)$, which represent the number of times each reaction has occurred in the time interval $[0,t)$. If a population contains $S_0$ and $I_0$ initial susceptibles and infectives, then $Z_1$ can occur a total of $S_0$ possible times, and $Z_2$ can occur up to $S_0 + I_0$ times. The DA process is preferred over the so-called ``population process" of counting population levels since the population levels at a given time can always be determined from the number of reactions that have occurred, but not vice versa. Specifically, \be \mathbf{X}(t) = \mathbf{X}(0) + \mathds{S}\mathbf{Z}(t), \ee
where $\mathbf{X}$ is a column vector with population levels of all population classes, $\mathbf{Z}$ is a column vector counting occurrences of all reactions, and $\mathds{S}$ is a matrix composed of the stoichiometric coefficients from Eq.\@ \ref{eq:reac}.
The components of $\vec{P}$ are ordered lexicographically, resulting in a lower triangular generator matrix $\mathds{A}$, since the reactions of infection and recovery cannot be reversed. Furthermore, due to the abundance of inaccessible states, $\mathds{A}$ is a sparse matrix.

To numerically integrate the master equation, we use an algorithm recently devised by Jenkinson and Goutsias \cite{jg} known as the \textit{implicit-Euler}, or IE, method. The IE method takes advantage of the sparsity of $\mathds{A}$ to achieve computational efficiency. We first discretize time into integer multiples of a time step $\tau$. Given specified initial conditions $\vec{P}(0)$, we can recursively solve the system of linear equations
\be (\mathds{1} - \tau \mathds{A}) \vec{P}(t_i) =  \vec{P}(t_{i-1}). \ee 
This is implemented in MATLAB.

In addition to the IE method, the stochastic SIR model is also frequently applied in direct Monte Carlo simulations \cite{fm1} \cite{fm2}, of which many repeated runs must be performed to obtain a probability distribution. In contrast, solving the master equation generates a full probability distribution at once for a chosen set of initial parameters. Moreover, it is also possible to arrive at the steady-state probability distribution without the need for iterative integration of the master equation \cite{houseetal} \cite{blackross}. However, because we simulate real-time vaccination, we need to obtain the probability vector at intermediate points during the epidemic, before the final steady state is reached. 

While the IE method can generate the full probability distribution at every time step with high temporal resolution, it requires much more computation time than solving a set of deterministic equations, realizing a direct Monte Carlo simulation, or acquiring the final steady-state probability distribution. Hence, we consider relatively small system sizes in this paper, especially in the case of two coupled populations, where the number of reactions doubles compared to a system of one individual population.

In individual populations, the number of possible states, and thus the dimension of the generator matrix, is (S(0) + I(0) + 1)(S(0) + 1). Thus, for $I(0) << S(0)$, the matrix dimension approximately scales as $N^2$. For multiple interacting populations, the number of possible states becomes 
\be \prod_i^M (S_i(0)+I_i(0)+1)(S_i(0)+1), \nn \ee where $M$ is the total number of populations. Therefore, for two coupled populations, the matrix dimension scales as $N^4$.

While the two coupled populations in our simulations may be small, they can represent realistic communities such as classrooms, groups of households or families, or small neighborhoods. Furthermore, the critical effects of temporal variability and non-transmission present in stochastic but not deterministic models also occur for much larger populations, which can represent entire cities or nations.

We initialize our simulations with specified numbers of susceptibles $S_0$ and infectives $I_0$, the reproductive number $r_0$, and the recovery rate $\gamma$. The infection rate $\beta$ is calculated from $r_0$ and $\gamma$. At $t=0$, the system is in state $(0,0)$ with probability 1; that is, the first element of $\vec{P}(0)$ is 1 with all others zero. 

The severity of the epidemic is quantified by the mean final epidemic size $\left<E\right>$ \cite{edwin}, defined as the total population minus the mean number of susceptibles remaining at the end of the epidemic. This represents the number of people who have been infected and then recovered, as measured at in the limit of infinite time:
\be \left<E\right> = \lim_{t \rightarrow \infty} N - \left<S(t)\right>. \ee
For our parameters, a simulation end time of 200 days was a sufficient approximation. In the deterministic model, the final epidemic size is exact, as is the number of remaining susceptibles.

\begin{figure}[h!] 
\includegraphics[width=.85\textwidth]{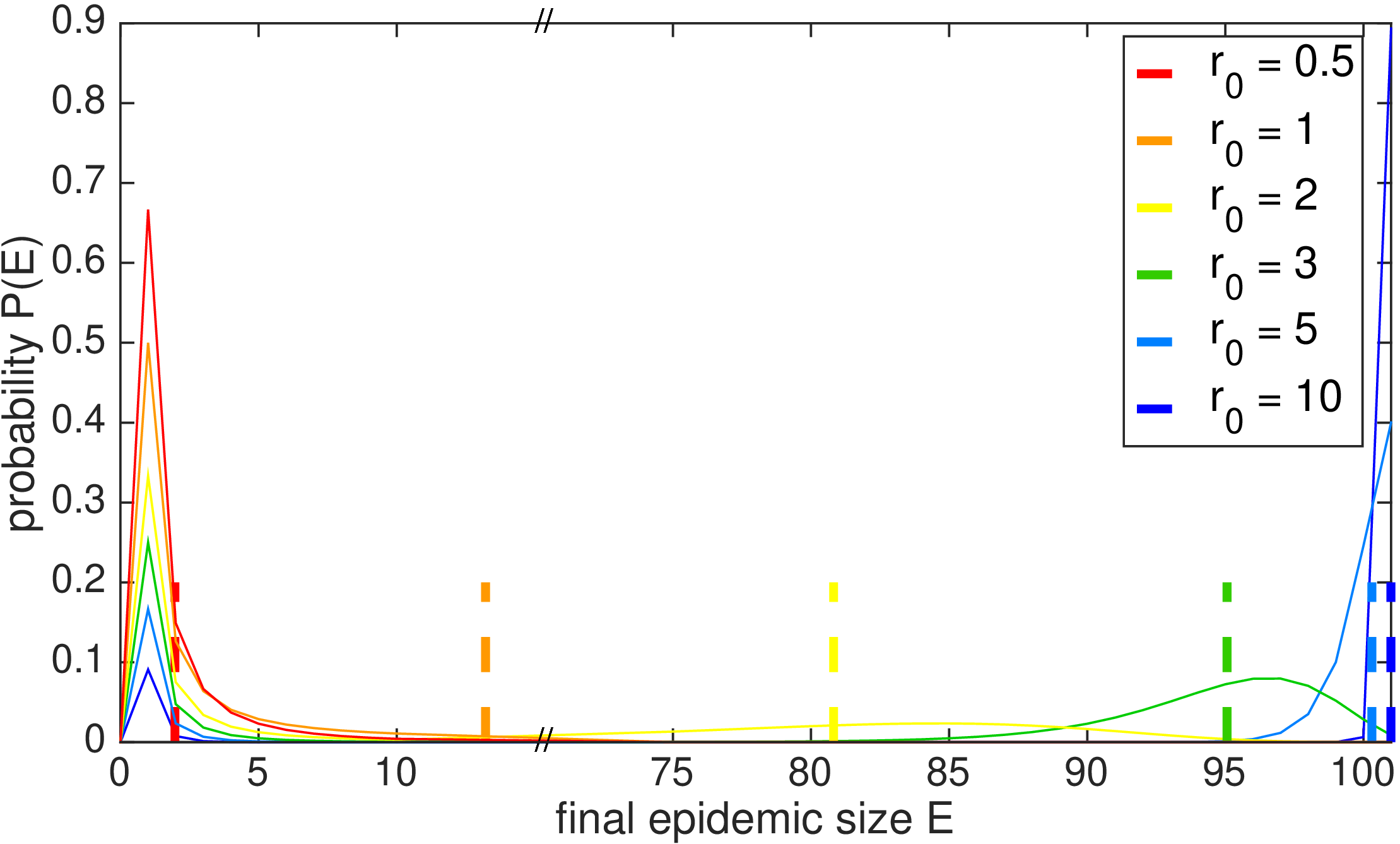}
\caption{{\bf Probability distribution of final epidemic size.} The distribution $P(E)$ is plotted for a population of 100 initial susceptibles and 1 infective in the stochastic model. The recovery rate $\gamma$ is set to $0.15$ and the reproductive number $r_0$ is varied between 0.5 and 10. The corresponding deterministic outcomes are represented by dashed lines.}
\label{fig:finalsize}
\end{figure}

The probability distribution $P(E)$ of final epidemic sizes $E$ (Fig.\@ \ref{fig:finalsize}) is observed to be bimodal for large $N$ and reproductive number $r_0 > 1$, as is characteristic of stochastic epidemic models \cite{bailey}: either a few individuals or a majority of the population become infected in the most likely scenarios. Fig.\@ \ref{fig:finalsize} illustrates the probability distribution for a simulation of one population containing 100 initial susceptibles, 1 initial infective, where the recovery rate $\gamma = 0.15$ and the reproductive number $r_0$ is varied between $0.5$ and $10$. The leftmost peaks are located at 1, the number of initial seed infectives, and represent the case in which very few individuals beyond the seed infective are infected, while the majority of the population remains susceptible. We denote this the \textit{terminal infection} case \cite{edwin}. The rightmost peaks represent the \textit{large-scale epidemic} case, in which the infection spreads throughout a large portion of the population. The deterministic outcomes are plotted as dashed lines and fall slightly to the left of the centers of the epidemic peaks. 
At large values of $r_0$, the large-scale epidemic peaks become taller and sharper; as $r_0$ decreases, the tails broaden, and the peaks are smaller. The terminal infection peaks are taller for lower values of $r_0$, since the probability of terminal infection for a fully susceptible population seeded with one infective is equal to $1/r_0$ \cite{kr}. For large $N$, when $r_0$ approaches unity, the distinction between the terminal infection and large-scale epidemic peaks disappear. 

\subsection*{Coupled Populations}
Due to the widespread network of transportation between cities, it is difficult for an outbreak of infectious disease to remain contained within one geographic location. Studying epidemics in isolated communities can still provide important insights; for example, abundant historical data on epidemics in the relatively isolated Faroe Islands have contributed to a rich understanding of the spatial spread of measles, whooping cough, and mumps \cite{faroe}. However, in this paper, we focus on the dependence of the severity of infection and the optimal vaccination protocol on the degree of interaction between cities.

We model the epidemic as occurring in two cities connected via transportation routes that allow individuals from one city to interact with those in the other, but where the total number of residents in each city does not change. This can be thought of as a system of commuters, where the degree of interaction, or coupling, between the cities is proportional to the time a commuter spends outside their home city. We consider the specific case in which an epidemic breaks out in one city and spreads to another. We seed one city, denoted city A, with one seed infective, while the other city, city B, contains only susceptibles.

For two coupled cities, there are now a total of four possible reactions, denoted $Z_1, Z_2, Z_3,$ and $Z_4$:

\be
\bad{
 Z_1 &: (S_A, I_A, S_B, I_B) \rightarrow (S_A -1, I_A + 1, S_B, I_B),\\ 
 Z_2 &: (S_A, I_A, S_B, I_B) \rightarrow (S_A, I_A, S_B-1, I_B+1),\\ 
 Z_3 &: (S_A, I_A, S_B, I_B) \rightarrow (S_A, I_A - 1, S_B, I_B),\\ 
 Z_4 &: (S_A, I_A, S_B, I_B) \rightarrow (S_A, I_A, S_B, I_B-1).} 
\label{eq:z1234}
\ee
A susceptible in one city can be infected by someone in either city; once infected, the susceptible becomes an infective of its own city.

In the stochastic model, the probability vector $\vec{P}$ consequently comprises probabilities of cumulative occurrences of the reactions in Eq.\@ \ref{eq:z1234},  $(Z_1, Z_2, Z_3, Z_4)$, ordered lexicographically. Our model also assumes homogeneous mixing within each city; that is, an individual in one city has the same probability of interacting with another individual in the same city, and the same probability of interacting with all individuals in the other city. While we abstract geographic details in our model, a distance-dependent kernel can be introduced to construct a more realistic model. That is, a function that decays with distance can be multiplied with the infection rate $\beta$, so that two individuals with greater spatial separation will be less likely to interact.
 
The coupling between the cities is characterized by a symmetric $2 \times 2$ matrix $f$, where $f_{ij}$ is the fraction of contacts an individual in city $i$ makes that are residents of city $j$ \cite{keelingshattock}. Each row and column of $f$ sums to 1. Thus, we denote the coupling between city A and city B as $f_{AB} = f_{BA}$. The rate of infection $\beta$ then becomes a $2 \times 2$ matrix with elements
\be \beta_{ij} = \frac{cp f_{ij}}{N_j}. \ee 

The corresponding deterministic equations that describe this model are
\ba{\deriv{S_A}{t} &= - \beta_{AA} S_A I_A - \beta_{AB}S_A I_B, \\ 
\deriv{S_B}{t} &= - \beta_{BA} S_B I_A - \beta_{BB}S_B I_B, \\ 
\deriv{I_A}{t}  &= \beta_{AA} S_A I_A +\beta_{AB}S_A I_B - \gamma I_A, \\
\deriv{I_B}{t}  &= \beta_{BA} S_B I_A +\beta_{BB}S_B I_B - \gamma I_B.}

If coupling between the two cities is low, it is possible that the infection does not spread at all from city A to city B in the stochastic model. The probability of an epidemic in city B is given by \cite{kr}:
\be \begin{split} \mathds{P}&(\text{infection spreading from A to B}) = \\ &1 - \exp{\left(-\beta_{BA} \left[1 - \frac{\gamma}{(\beta_{BA}+\beta_{BB})}\right] \int_0^{\infty} I_A(s) ds\right)} < 1 - \exp (-\beta_{BA}/\gamma).\end{split} \ee

In both models, there exists a lag between the peaks of the infections (i.e. where the number of infectives as a function of time $I(t)$ reaches a maximum) in the two cities. In the stochastic model, the lag time has a high variability, but on average is larger than the deterministic value due to the decreased probability of transmission.
In the deterministic model, if the two cities have nonzero coupling, then the infection is present in city B at the start, and the lag time is generally shorter. Unlike in the stochastic model, discrete integer values of infectives are not required to make contact in order for infection to spread. In the early stages of infection, when the number of susceptibles in city B is large and nearly constant, the number of infectives in city B grows exponentially at a rate $\beta_{BB} - \gamma$ \cite{kr}.

\subsection*{Vaccination}
One major objective of vaccination is to approach ``herd immunity", which occurs when a critical fraction of the population is vaccinated such that the effective reproductive number $r_{\text{eff}}$ drops below 1, thus preventing the epidemic from growing. If $V$ individuals are vaccinated, the effective reproductive number is $r_{\text{eff}} = r_0 (1-V/S_0) $.
Hence, the minimum number of individuals that must be vaccinated to induce herd immunity is $S_0(1-1/r_0)$.

Previous studies have utilized stochastic models to optimize vaccine allocation, but have largely focused on prophylaxis \cite{edwin} \cite{blackhouse}, rather than real-time vaccination, to which primarily deterministic models have been applied \cite{flu} \cite{birdflu} \cite{optdet} \cite{klepac}. 
In this paper, we determine the optimal allocation of vaccine, delivered in real time, for varying delay periods, coupling, and amounts of available vaccine. 

The optimal protocol is determined by minimizing the expected final epidemic size with respect to the fraction of vaccine allocated to each city. Only susceptibles are vaccinated in our simulations. The vaccines have 100\% efficacy and are delivered in discrete amounts. Furthermore, vaccines are released all at once on a specified time step. At this vaccination time step, the generator matrix is modified so that the transition rates reflect the new reduced number of susceptibles. First, a sparse generator matrix for a system containing the new population levels is computed and mapped to the dimensions of the original matrix. However, this does not account for the probability that the system is in certain states that, while previously accessible, are rendered newly inaccessible upon vaccination. For instance, consider a population initially containing two susceptibles and one infective. If allowed to evolve without intervention, it is possible for the two susceptibles to become infected, such that there are now three infectives. That is, the probability vector $\vec{P}$ is non-zero for the $Z_1 = 2, Z_2 = 0$ state. If we vaccinate both susceptibles at some time step, the system cannot transition to a state of higher infection, but should be allowed to transition out of this $(2,0)$ state through recovery, i.e. to states $(2,0)$, $(2,1)$, $(2,2)$, and $(2,3)$. Therefore, terms are added to the new generator matrix to reflect this. 

\section*{Results}Time delays in reactive vaccination are inevitable. Factors such as logistics, spatial separation, and limited availability of vaccine prevent the populace from receiving vaccination immediately upon the outbreak of infection. Vaccine stockpiles can mitigate but not fully eliminate this time delay. Furthermore, delayed efficacy of vaccines can also contribute to a delay between outbreak and immunity of vaccinated susceptibles.
Within an individual population, when time delay is low, the amount of vaccine required to keep the average epidemic size below a bounded value increases roughly exponentially with time delay. However, for large enough time delay, any increase in vaccine will not affect the final epidemic size, since the infection will have already spread to the entire population.

When individuals from two populations interact, coupling also contributes to tradeoffs, along with the amount of available vaccine and the time delay. In general, when coupling is very low, the optimal strategy tends to disparately favor one city over another, an effect that is less prominent as time delay increases. If coupling is high, the cities are more well-mixed, and the optimal protocol usually allocates vaccine to each city in proportional amounts.
 
We first investigate the spread of an arbitrary disease with $r_0 = 2$ and $\gamma = 0.15$ through an individual, non-interacting population of 100 initial susceptibles and one initial infective. With spatial, logistical, and technical details abstracted, we identify tradeoffs between the amount of available vaccine and the time delay until vaccine is administered. We then examine the spread of this disease from the host city, city A, to an initially infection-free city, city B. Due to the computational intensity of the IE method, we focus on a system of two cities of 40 individuals each, all susceptible except for one seed infective in city A. We then investigate how coupling, as well as vaccine and time delay, contributes to tradeoffs, and we identify the optimal vaccine allocation protocols which minimize the final epidemic size. We also examine the ``worst-case scenarios'' which result when the final epidemic size is maximized, since common epidemic control strategies, such as prophylaxis, can result in a mathematically non-optimal scenario. We compare the optimal strategies with the worst-case protocols to determine where intervention has the most significant payoffs.

\subsection*{Tradeoffs between Time Delay and Available Vaccine} 
\begin{figure}[h!]
\subfloat{\includegraphics[width=0.49\textwidth]{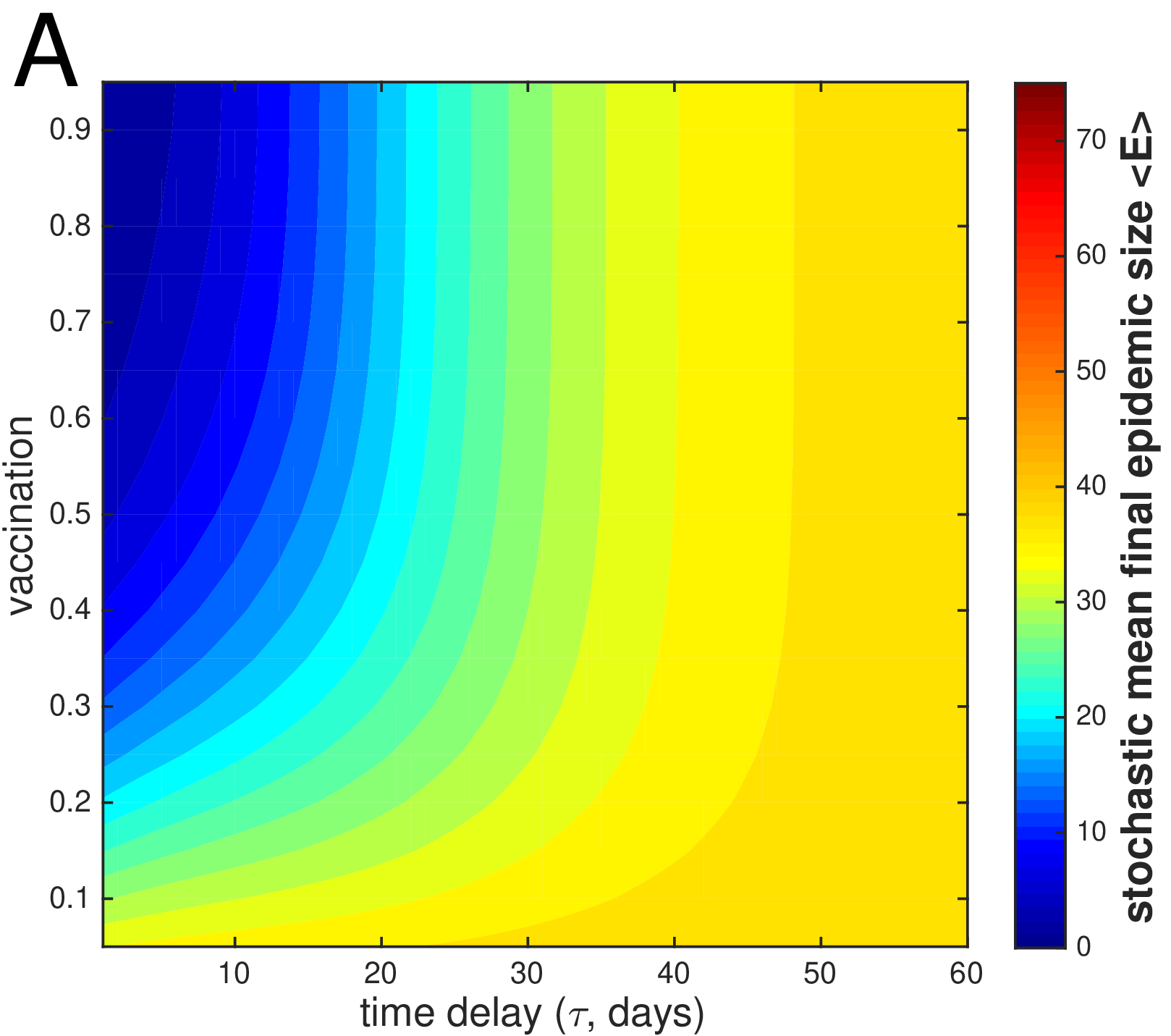}
\label{tdstoch}} \ \ \
\subfloat{\includegraphics[width=0.49\textwidth]{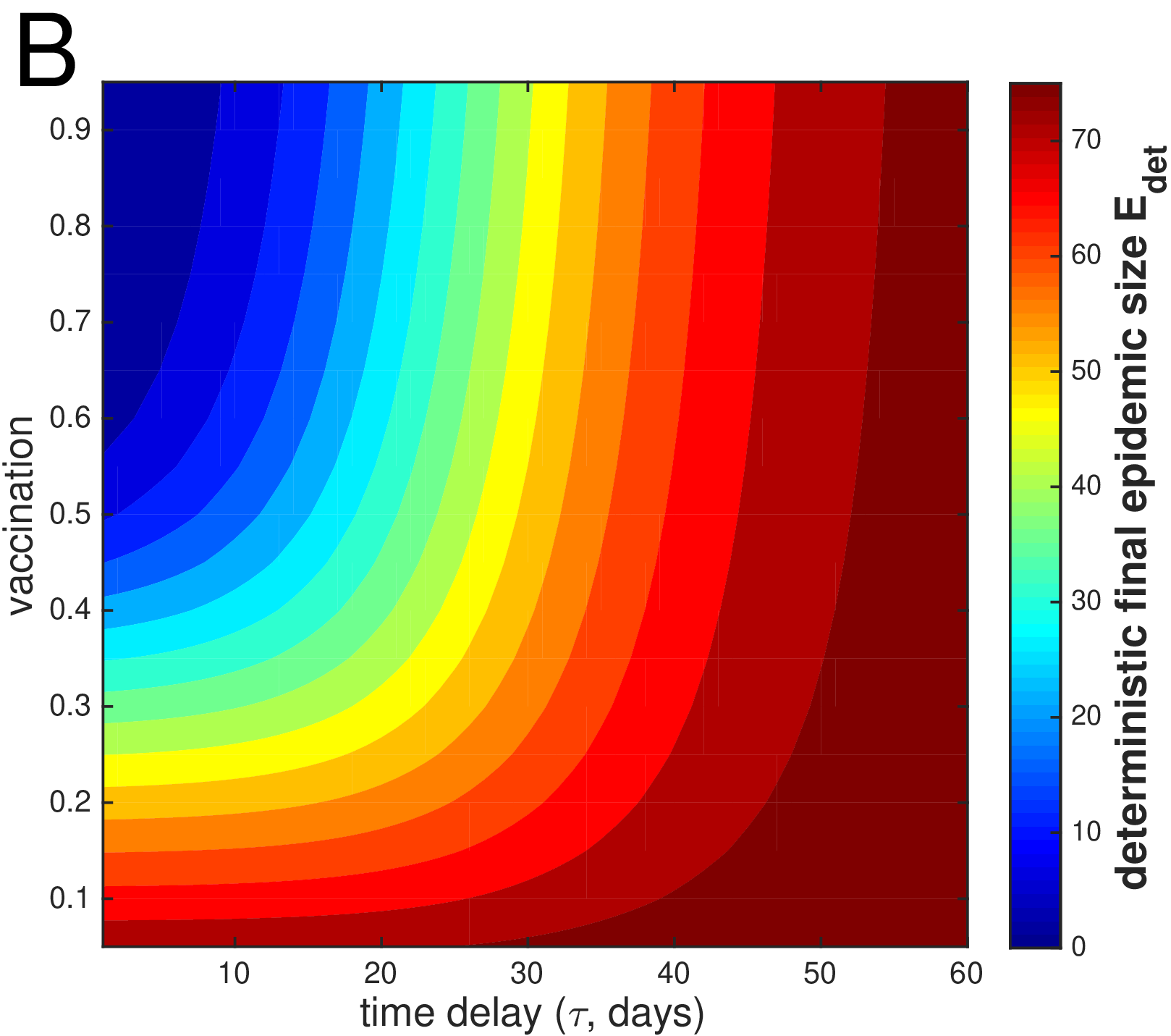}
\label{tddet}}
\caption{{\bf Tradeoffs between time delay and available vaccine represented by contours of final epidemic size.} The mean final epidemic size in the stochastic model $\left<E\right>$ (panel A) and the exact final epidemic size in the deterministic model $E_{\text{det}}$ (panel B) are plotted as a function of time delay $\tau$ (in days) and fraction of total population vaccinated. Both models contain 100 initial susceptibles, 1 initial infective, $\gamma = 0.15$, and $r_0 = 2$.}
\label{fig:timedelay}
\end{figure}

The tradeoff between time delay, denoted $\tau$, and available vaccine in a single population is illustrated in Fig.\@ \ref{fig:timedelay}, which shows contours of final epidemic size plotted as a function of vaccination and time delay for both the stochastic model (Fig.\@ \ref{tdstoch}) and the deterministic model (Fig.\@ \ref{tddet}). The amount of vaccine is expressed as the fraction of the total population which is vaccinated. The simulation is run on a population of 100 initial susceptibles and 1 seed infective, with reproductive number $r_0 = 2.0$ and recovery rate $\gamma = 0.15$. The deterministic results approach a final epidemic size of approximately 70 individuals at day 50, compared to a mean final size of approximately 35 individuals in the stochastic model. The epidemic is more severe in the deterministic model than in the stochastic model since the deterministic model does not require discrete integer values of individuals to make contact, thereby facilitating the infectious spread. 

However, the contours in the two plots are similarly shaped. For both models, early in the epidemic, a smaller amount of vaccine is sufficient to contain the outbreak, but as time delay increases, more and more vaccine is required to prevent the epidemic from growing beyond a certain size. For $\tau$ below about 20 days, the contours are approximately linear in semi-logarithmic space; that is, the vaccination rate must increase roughly  exponentially with time delay to keep the final epidemic size bounded below a certain value. Eventually, if time delay is great enough, any increase in the amount of vaccination will be ineffective in reducing epidemic size, as evidenced by the rapidly increasing, nearly vertical contours. This transition between regimes is more abrupt in the stochastic model, for which the contours sharply become essentially vertical at late time delay.

\begin{figure}[h!] 
\includegraphics[width=0.75\textwidth]{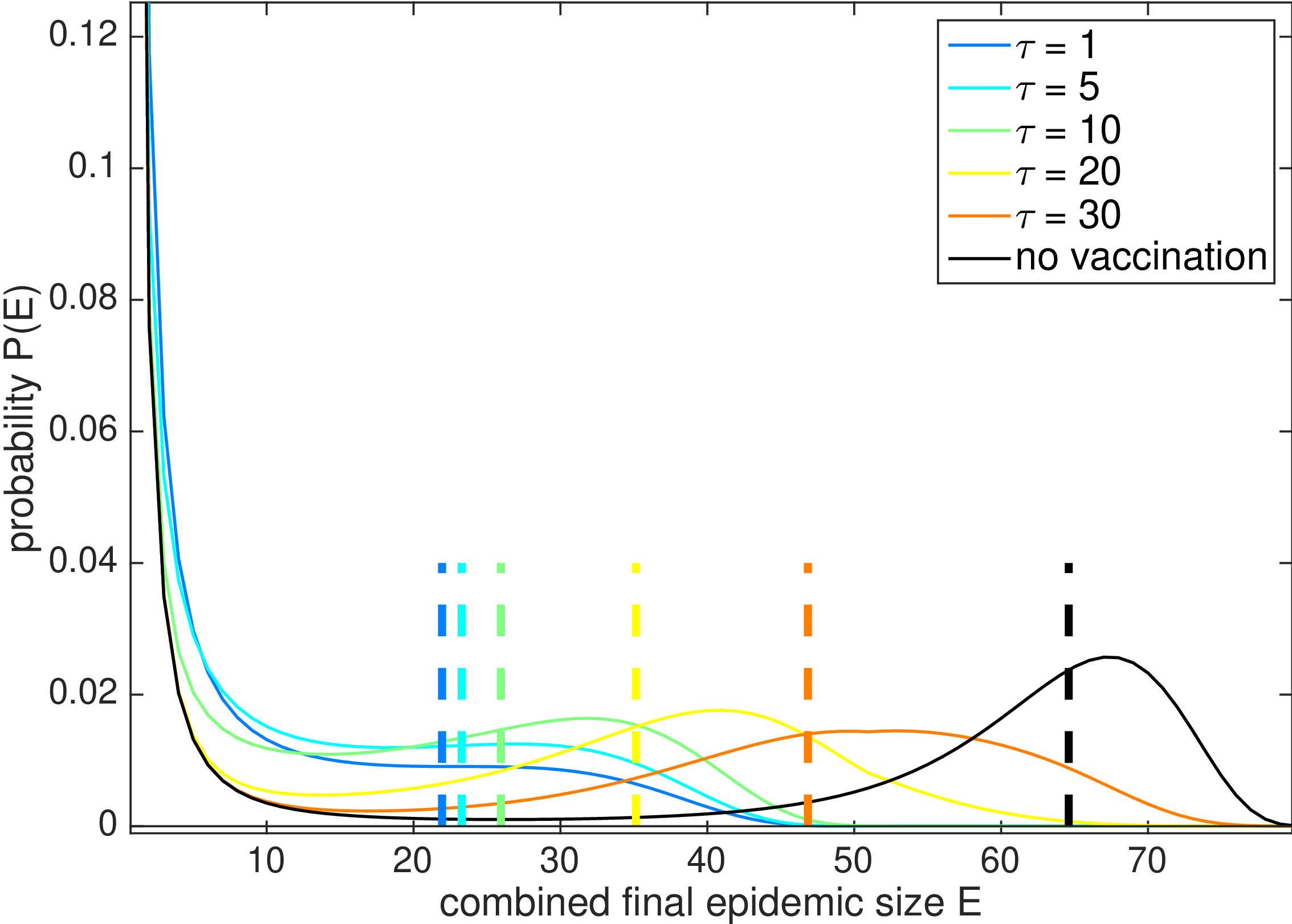}
\caption{{\bf Probability distribution of combined final epidemic size in two coupled populations.} The probability distribution of $E = E_A + E_B$ is plotted for two coupled populations of 40 people each, with one initial infective in city A. The cities have coupling $f_{AB} = 0.25$ and are given 15 vaccines each ($37.5\%$ of the population) at varying values of time delay $\tau$. The chosen parameter values are $r_0 = 2$ and $\gamma = 0.15$.}
\label{fig:moo}
\end{figure}

To further illustrate the increase in epidemic size that results from increased time delay, Fig.\@ \ref{fig:moo} plots the probability distributions of combined final epidemic size $E = E_A + E_B$ for two coupled populations of 40 people each. City A has 39 initial susceptibles and one initial infective, while city B is entirely susceptible. Five simulations are run with 15 susceptibles in each city vaccinated at time delay $\tau = $ 1, 5, 10, 20, and 30 days, respectively. The cities have the same coupling, $\fab = 0.25$, across all simulations. If vaccination were completely successful, the final epidemic size should be at most 50. The probability of $E > 50$ is approximately zero if the cities are vaccinated by 10 days, indicating successful vaccination. However, at a time delay of 20 days or later, the large-scale epidemic peaks broaden, and $P(E > 50)$ is nonzero. At $\tau = 20$ days, the tail of the large-scale epidemic peak in the probability distribution terminates at $E \approx 70$. This indicates that the disease may have spread to more than 50 individuals by the vaccination date, leaving fewer than 30 susceptibles remaining and thereby resulting in potentially wasted vaccine.

\subsection*{Tradeoffs Involving Time Delay, Available Vaccine, and Coupling}
The level of interaction between individuals living in two different cities can affect the severity of the combined epidemic size and the speed at which the infection spreads from the host city to the other. The coupling $\fab$ between city A and city B might be determined by spatial separation and/or frequency of mass transportation. We now examine the affect of coupling $\fab$ and the ratio of vaccine allocated to each city, in addition to time delay $\tau$ and amount of vaccine, on the final epidemic size.

\begin{figure}[h!]
\subfloat{\includegraphics[width=0.49\textwidth]{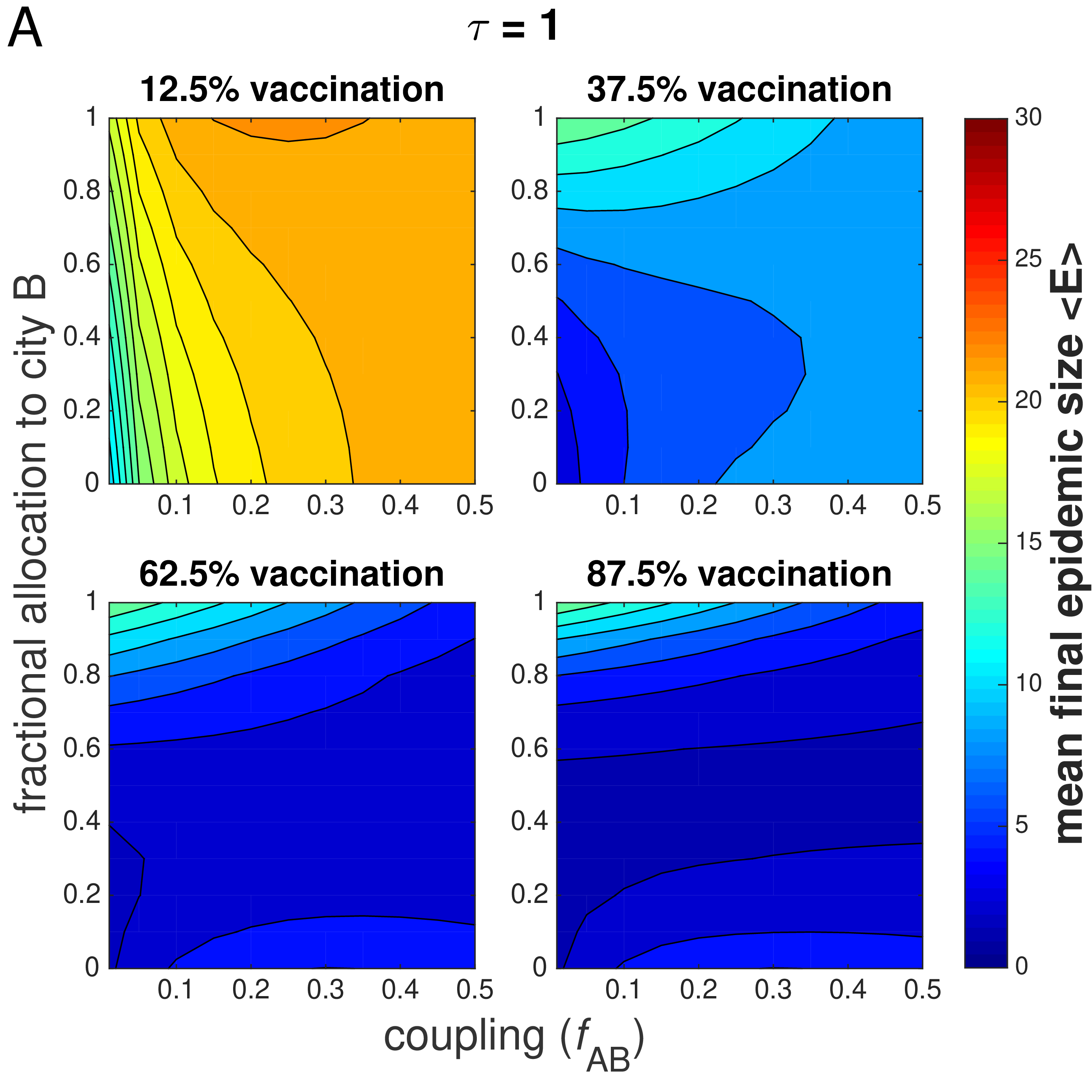}
\label{tau1}} \ \ \
\subfloat{\includegraphics[width=0.49\textwidth]{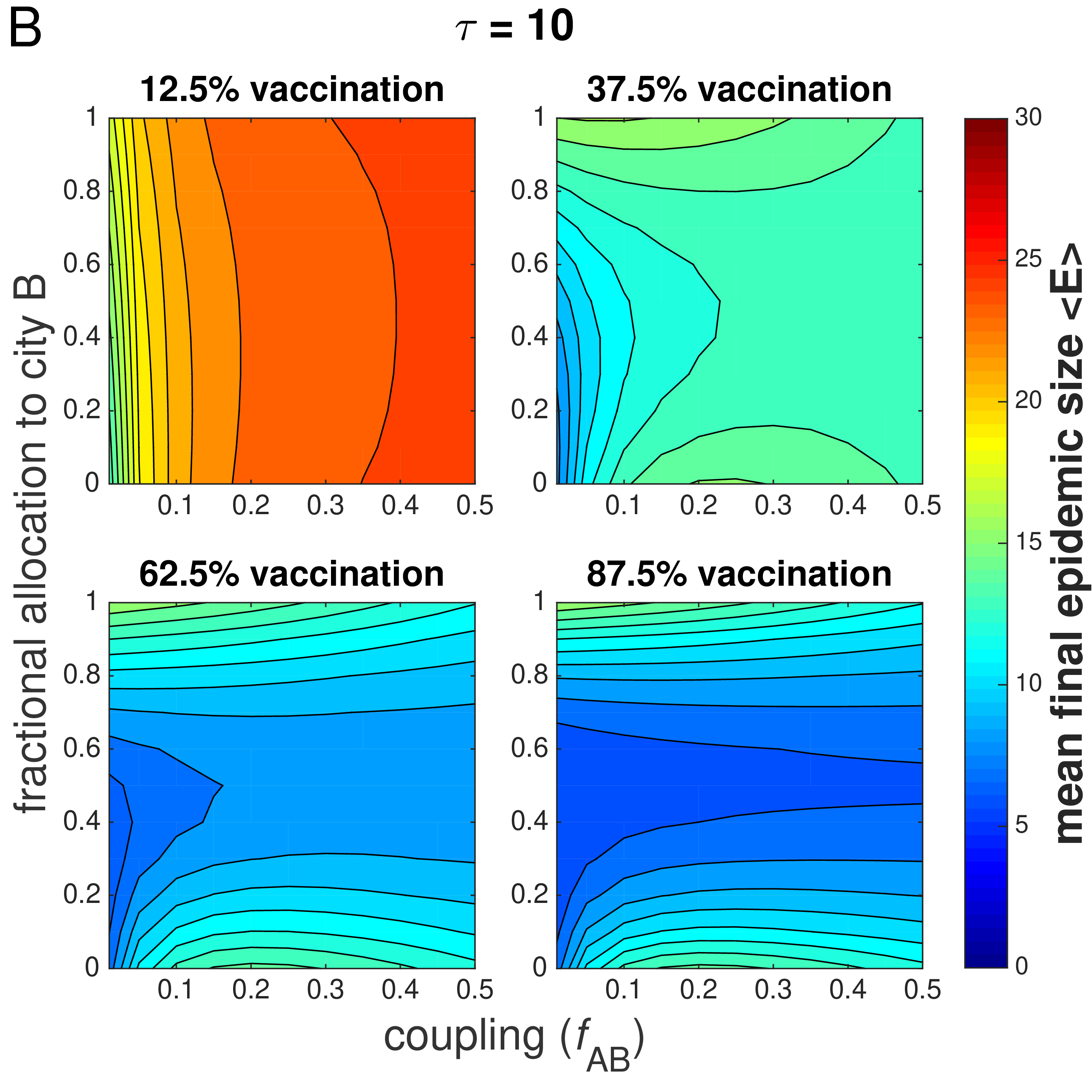}
\label{tau2}}
\caption{{\bf Tradeoffs involving coupling, time delay and available vaccine represented by contours of combined final epidemic size.} The mean final epidemic size for two coupled cities $\left<E\right>$ is plotted as a function of fractional vaccine allocation to city B and coupling $\fab$ for varying amounts of available vaccine at $\tau = 1$ day (panel A) and $\tau = 10$ days (panel B). City A has 39 initial susceptibles and 1 infective; city B has 40 initial susceptibles. The recovery rate $\gamma = 0.15$ and the reproductive number $r_0 = 2$.}
\label{fig:t1}
\end{figure}

The mean final epidemic sizes at two different values of time delay, $\tau = 1$ day and $\tau = 10$ days, are compared in Figs.\@ \ref{tau1} and \ref{tau2}, which plot the mean final size $\left<E\right>$ as a function of the fractional vaccine allocation to city B and the coupling $\fab$. The color scale is kept the same across all panels.
The amount of available resources does not correspond to the actual fraction of the population that is vaccinated, since for certain allocation protocols, some vaccines might be wasted. For example, if there are 60 vaccines, enough to vaccinate 75\% of the combined population, allocating all available vaccine to city B, which has a population of 40,  would result in 20 wasted vaccines. 

Vertical contours arise in the limiting case when, for a certain coupling, the final epidemic size does not depend on the allocation of vaccine. On the other hand, horizontal contours arise in the limit where the final epidemic size depends strongly on the vaccine allocation, but not on the coupling.

For $\tau = 1$, in the first panel of Fig.\@ \ref{tau1}, there are 10 available vaccines, or enough resources to vaccinate at most $12.5$\% of the population.
In this case, the mean final epidemic size is smallest when a majority of resources is allocated to city A, and when coupling is low. The largest possible mean final size of $\left<E\right> = 22$ occurs when all vaccine is allocated to city B for moderate coupling. This indicates that prophylaxis (i.e., preemptively vaccinating an uninfected population) is not the best course of action; rather, very shortly after an outbreak, it is optimal to deliver all available resources to the city where the outbreak has occurred, especially if the cities are weakly coupled, in order to increase the chances of preventing the epidemic from spreading to city B.

In the second panel, when 30 vaccines are available (enough for $37.5$\% of the population), the contours are asymmetrical about the horizontal. For low $\fab$, the epidemic size is minimized when most vaccine is given to city A, but as $\fab$ increases, the minima occur for an increasingly equal allocation. 
For larger amounts of vaccine, the preferred course of action is to allocate equal amounts of vaccine to each city, except when $\fab$ is extremely low, in which case an equal allocation has roughly the same effect as one which allocates all vaccine to city A.

Contours of mean final epidemic size as a function of vaccine allocation and coupling for $\tau = 10$ days are plotted in Fig. \ref{tau2}. In general, the final epidemic sizes are greater with the increased time delay. When at most $12.5$\% of the total population can be vaccinated, the contours are roughly vertical. Again, the smallest epidemic size for these parameters will occur when all vaccine is given to city A for very low $\fab$, but for most values of $\fab$, the final size does not depend strongly on the allocation. In the remaining panels, the contours for larger amounts of vaccine are more symmetrical about the horizontal than the corresponding plots for $\tau = 1$ day, and more often favor an equal allocation, rather than a protocol which disparately allocates most vaccine to one city or another.

\subsection*{Optimal Vaccination Strategies}
\begin{figure}[h!]
\includegraphics[width=0.75\textwidth]{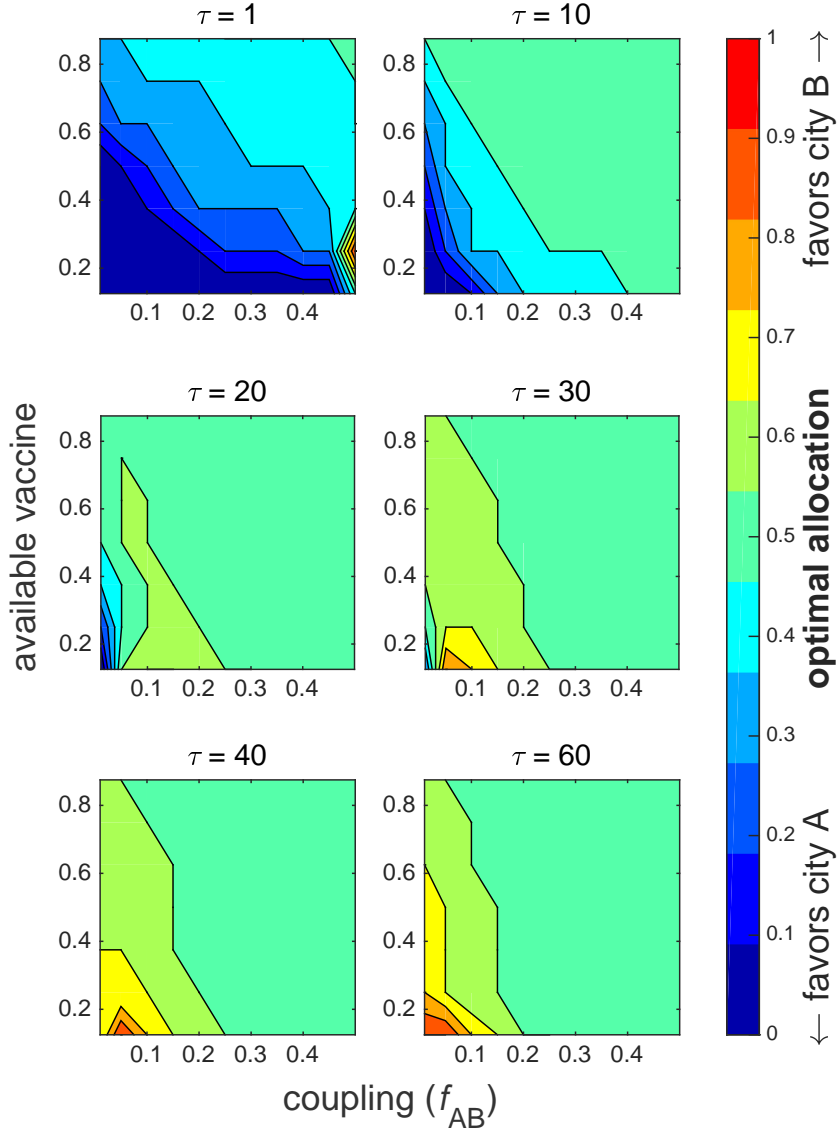}
\caption{{\bf Optimal vaccination strategies at different time delays.} The optimal fraction of total vaccine allocated to city B is plotted as a function of available vaccine (expressed as a fraction of the total combined population) and coupling $\fab$ for different fixed values of time delay $\tau$ ranging from 1 to 60 days. City A has 39 initial susceptibles and 1 infective; city B has 40 initial susceptibles. The recovery rate $\gamma = 0.15$ and the reproductive number $r_0 = 2$.}
\label{fig:opt1}
\end{figure}

Optimal protocols were determined by minimizing the mean final epidemic size $\left<E\right>$ in the stochastic model, and the final epidemic size $E_{\text{det}}$ in the deterministic model. For the stochastic model, Fig.\@ \ref{fig:opt1} plots all optimal allocations (expressed as the fraction allocated to city B) as a function of available vaccine and coupling $\fab$. Each subplot is a snapshot taken at increasing values of fixed time delay $\tau$. Again, the simulations are of two cities, each containing 40 total individuals, with one infected and the remaining susceptible in city A, and all susceptible in city B.

The first panel describes the optimal protocol very early after the outbreak, when there is a 1-day time delay. In this case, the optimal protocol allocates all resources to city A if the available vaccine is low, unless coupling $\fab$ is very high. In that event, a small fraction of the vaccine should also be given to city B, since the cities are more well-mixed. As time delay increases, completely favoring city A is optimal only for increasingly narrower ranges of weak coupling and low vaccine. The contours decrease with vaccine roughly linearly as a function of coupling. 

However, as $\tau$ increases to 20 days (the third panel), for moderately weak coupling $0.05 \lesssim \fab \lesssim 0.25$, the optimal protocol slightly favors city B, allocating 60\% of available vaccine, while equal allocation is preferred for most remaining scenarios. Since each city has approximately equal numbers of initial susceptibles, an equal allocation is the same as a proportional allocation. At $\tau > 40$ days, for weak coupling ($\fab < 0.1$), the optimal protocol favors city B, especially when vaccine is low, indicating that the epidemic has begun to die out in city A while it is growing in city B.

\begin{figure}[h!] 
\includegraphics[width=0.75\textwidth]{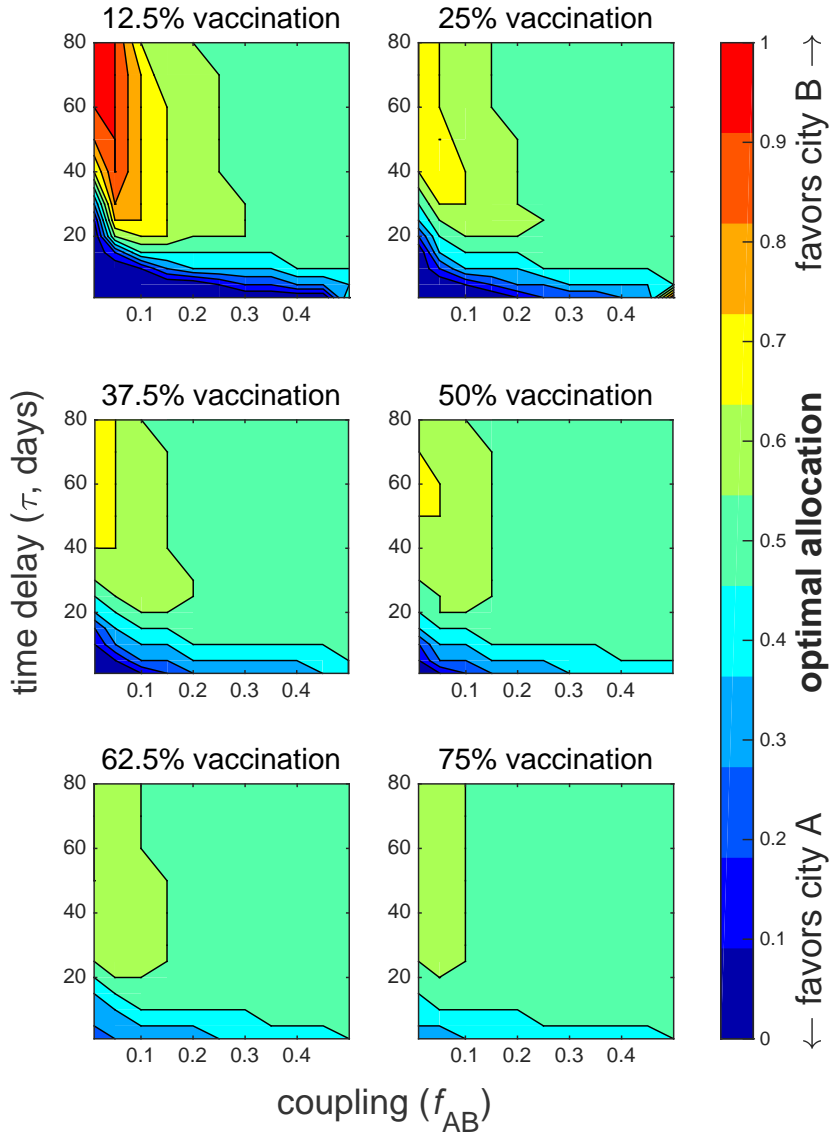}
\caption{{\bf Optimal vaccination strategies for different amounts of available vaccine.} The optimal fraction of total vaccine allocated to city B is plotted as a function of time delay $\tau$ and coupling $\fab$ for fixed vaccination amounts ranging from 10 to 60 vaccines. The remaining parameters are the same as in Fig. \ref{fig:opt1}.}
\label{fig:opt2}
\end{figure}

Fig.\@ \ref{fig:opt2} contains the same information, except plotted as a function of time delay and coupling, with snapshots taken at different amounts of available vaccine. In the first panel, for 12.5\% vaccination, there is high variability in the optimal protocols. For weak coupling, the optimal protocol allocates almost all vaccine to city B at large time delay, indicating situations when the epidemic has weakened considerably in city A and has spread to city B. However, for small $\tau \lesssim 20$, the optimal protocol allocates all vaccine to city A, regardless of coupling. At large coupling, the optimal protocol is an equal allocation for $\tau > 20$.

For greater amounts of vaccine, the contours remain a similar shape. As the available amount of vaccine increases, the optimal protocols increasingly approach an equal allocation. In the last panel, for 75\% vaccination, the optimal protocol allocates vaccine equally for almost all $\tau$ and $\fab$. The exceptions are a 30-40\% allocation of vaccine to city B for small $\tau$, and 60\% allocation of vaccine to city B for large $\tau$ and $\fab < 0.1$. These allocations correspond to one city receiving 42 vaccines and the other receiving 28. Since there are only 40 individuals in each city, at least 2 vaccines are wasted.

The deterministic model, in comparison, exhibits strikingly different results, as illustrated in Figs.\@ S1 and S2. Fig.\@ S1 should be compared with Fig.\@ \ref{fig:opt1}, and Fig.\@ S2 with Fig.\@ \ref{fig:opt2}. While for low $\tau$, the stochastic model tends to favor city A, the deterministic optimum is closer to an equal allocation for most values of coupling and vaccine.

\begin{figure}[h!]
\includegraphics[width=0.75\textwidth]{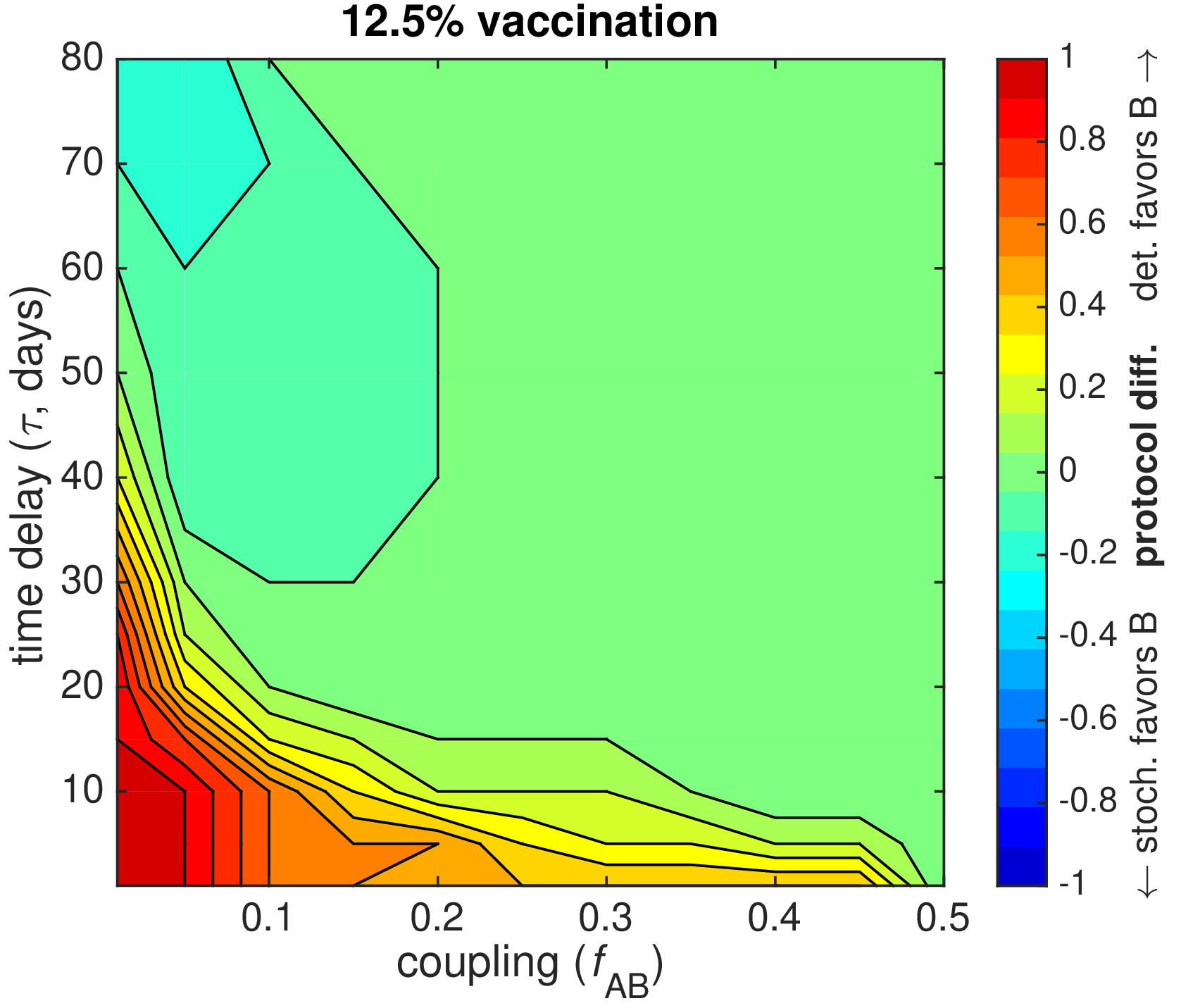}
\caption{{\bf Difference between stochastic and deterministic optimal strategies.} The stochastic and deterministic models are compared by plotting the difference between optimal fractions of vaccine allocated to city B for the case of 12.5\% vaccination. City A has 39 initial susceptibles and 1 infective; city B has 40 initial susceptibles. The recovery rate $\gamma = 0.15$ and the reproductive number $r_0 = 2$.}
\label{fig:diff}
\end{figure}

To directly compare the stochastic and deterministic models, we plot the difference between the optimal fractions allocated to city B prescribed by the two models in Fig.\@ \ref{fig:diff}. A positive difference indicates that the deterministic model favors city B more heavily than the stochastic model, and a negative difference indicates vice versa. We specifically highlight the case of 12.5\% vaccination, since lower vaccine leads to protocol differences further from zero.

In Fig.\@ \ref{fig:diff}, the difference is always greater than $-0.2$ and for the most part positive. The deterministic model generally allocates more vaccine to city B than the stochastic model, particularly for weak coupling and low vaccine. When $\tau$ and $\fab$ are small, the deterministic model completely favors city B, and the stochastic model completely favors city A, leading to a protocol difference of 1. When $\fab$ is large, the deterministic and stochastic models both favor an equal allocation, so that the difference is 0. However, at large $\tau > 30$ and small $\fab$, the difference is negative. In this case, the stochastic model slightly favors city B compared to the deterministic model, due to the asynchronous spread of infection and the longer temporal separation between epidemic peaks at low coupling.

In general, the deterministic model requires a larger amount of vaccination to eradicate the epidemic than in the stochastic model. In the deterministic model, the infection will always spread from city A to city B if coupling is nonzero, whereas in the stochastic model, the infection has a nonzero probability of terminating in city A. In the early stages of an epidemic, the stochastic optimal protocol tends to favor city A, in order to decrease the likelihood of inter-city transmission, rather than preemptively vaccinating city B. In contrast, the deterministic optimal protocol tends to favor city B, in order to ultimately minimize final epidemic size. Furthermore, in the stochastic model, as vaccination increases, the epidemic is more likely to be eradicated by stochastic extinction before the amount of vaccine required for herd immunity in a deterministic model is reached \cite{kr}.

\subsection*{Worst-Optimal Differences}
The worst-case vaccination protocol is defined as the vaccine allocation which results in the \textit{maximum} possible mean final epidemic size. In some cases, the mathematically worst-case scenario results from common epidemic control strategies such as prophylaxis, particularly in a region where the outbreak has not occurred, rather than sending vaccine to the active region in order to mitigate the outbreak.
The worst-case vaccination protocols are plotted for the stochastic model in Fig.\@ S3. The worst-case scenario will either result from allocating all vaccine to city A or to city B, depending primarily on the time delay.
When time delay is low ($\tau \lesssim 20$ days), the worst-case protocol allocates all available vaccine to city B. When time delay is high, the worst-case protocol allocates all vaccine to city A.

\begin{figure}[h!] 
\includegraphics[width=0.75\textwidth]{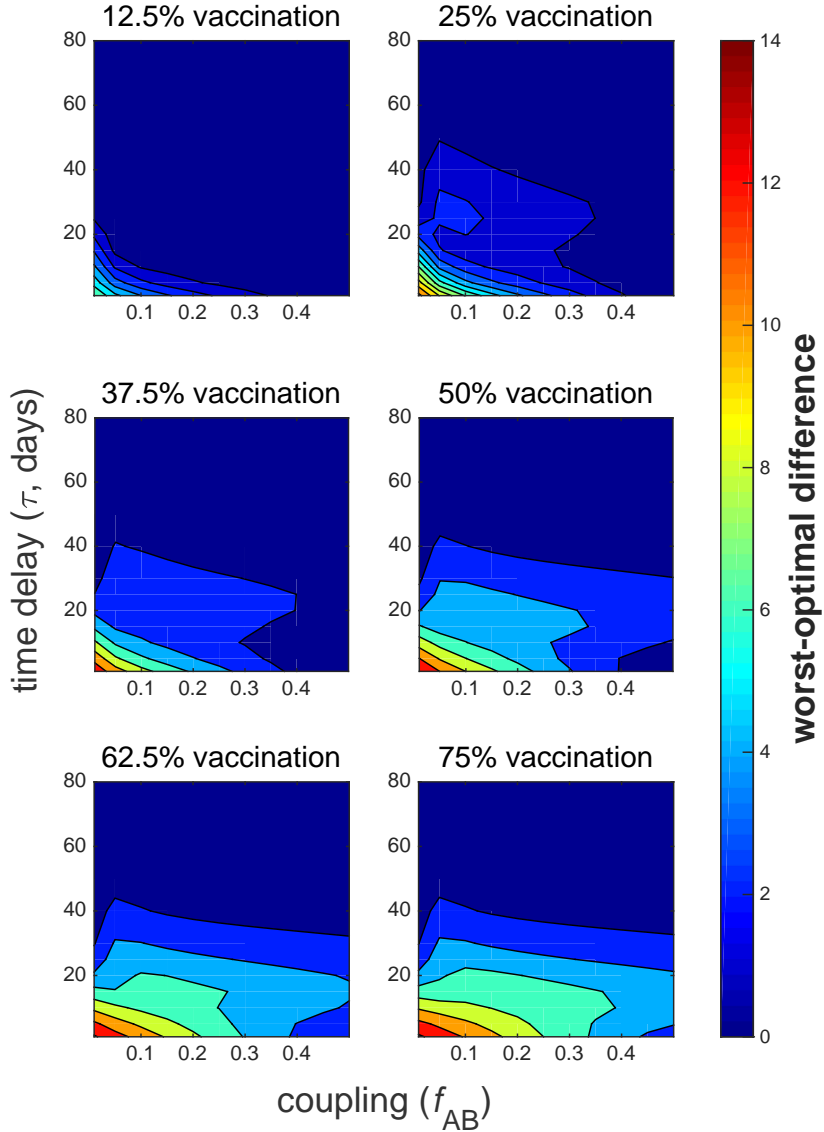}
\caption{{\bf Difference between optimal and worst-case final epidemic sizes.} The difference in stochastic mean final epidemic size between worst-case and optimal protocols is plotted as a function of time delay $\tau$ and coupling $f_{\text{AB}}$ for increasing amounts of available vaccine. The remaining parameters are the same as in Fig. \ref{fig:diff}.}
\label{fig:worstopt}
\end{figure}

The difference in mean final epidemic size $\left<E\right>$ between the optimal and worst-case scenarios, or the \textit{worst-optimal difference}, is plotted as a function of time delay and coupling for fixed values of available vaccine in Fig.\@ \ref{fig:worstopt}. 
The largest worst-optimal difference is about $\left<E\right> = 14$, or $17.5\%$ of the total combined population, and occurs for weak coupling, small time delay, and large amounts of vaccine. This implies that optimal vaccination is most effective when administered early in an epidemic, since the infection will not have significantly propagated. This case corresponds to an optimal allocation that tends to be relatively disparate in favoring one population over another. The worst-optimal difference decreases as time delay and coupling increases, both of which correspond to an enhanced spread of disease in city B. The worst-optimal difference is almost non-existent for small time delay in well-mixed populations, or for large time delay regardless of coupling. At large time delay, the epidemic has progressed to the point that the worst-optimal difference is almost zero; strategic vaccination, or even vaccination at all, will be mostly futile.

In contrast, the deterministic worst-case scenarios all result when vaccine is allocated entirely to city A. The contours of the deterministic worst-optimal plot take on similar shapes as in the stochastic plot, and the largest differences also occur for low time delay, low coupling, and high vaccine. However, the maximum difference is about twice that of the stochastic result, at about $E_{\text{det}} = 30$, or $37.5\%$ of the population, compared to $17.5\%$ in the stochastic model. 
This suggests that the timing of the vaccination is more important in the stochastic model than its allocation. The stochastic model is less sensitive to precise optimization of resource allocation, but the stochastic optimal protocols (Fig.\@ \ref{fig:opt2}) depend more strongly on time than the deterministic protocols (Fig.\@ S2). The deterministic worst-optimal differences are plotted in Fig.\@ S5, while the worst-case resource allocation protocols are shown in Fig.\@ S4.

\begin{figure}[h!] 
\subfloat{\includegraphics[width=0.49\textwidth]{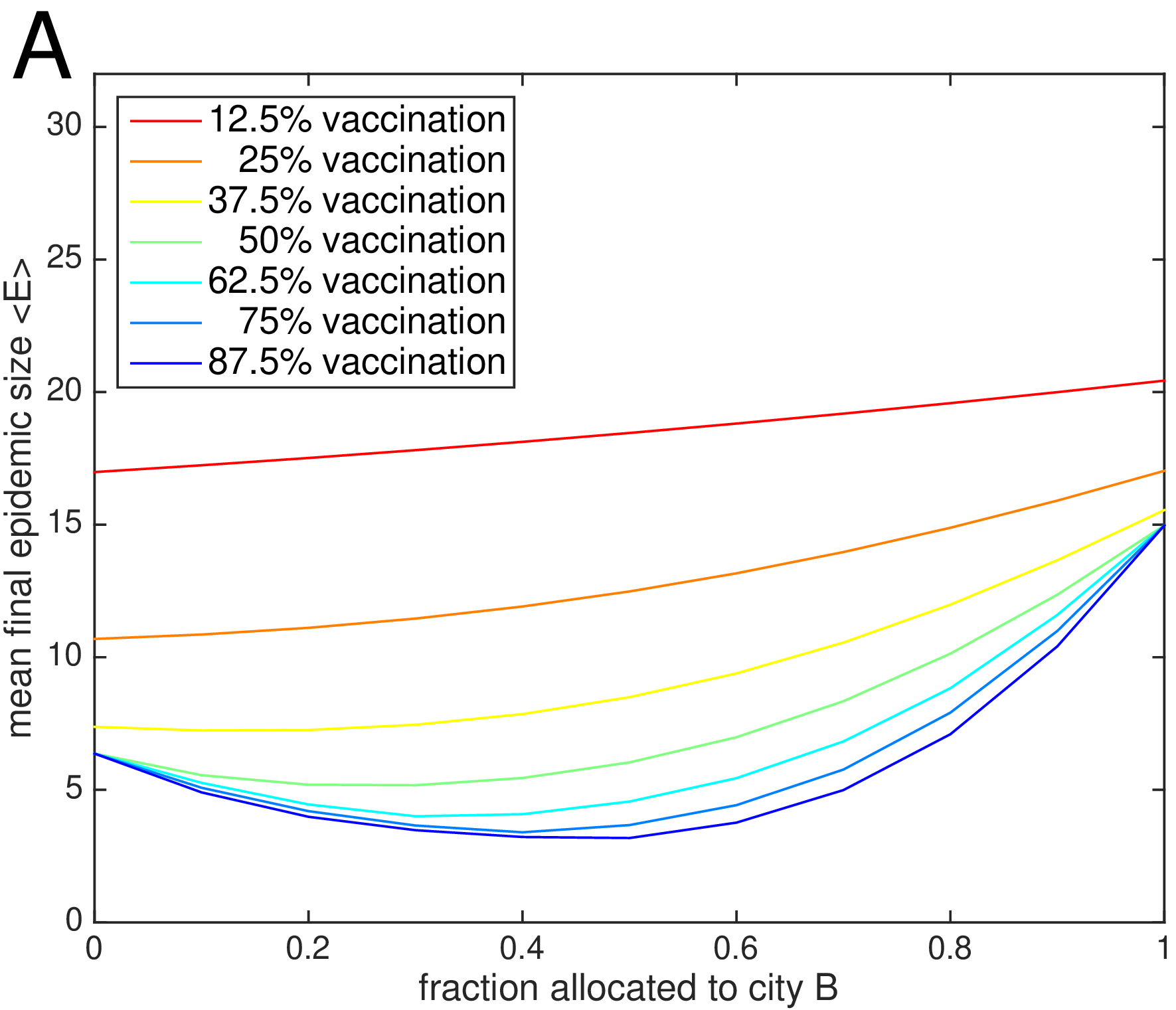}
\label{ex}} \ \ \
\subfloat{\includegraphics[width=0.49\textwidth]{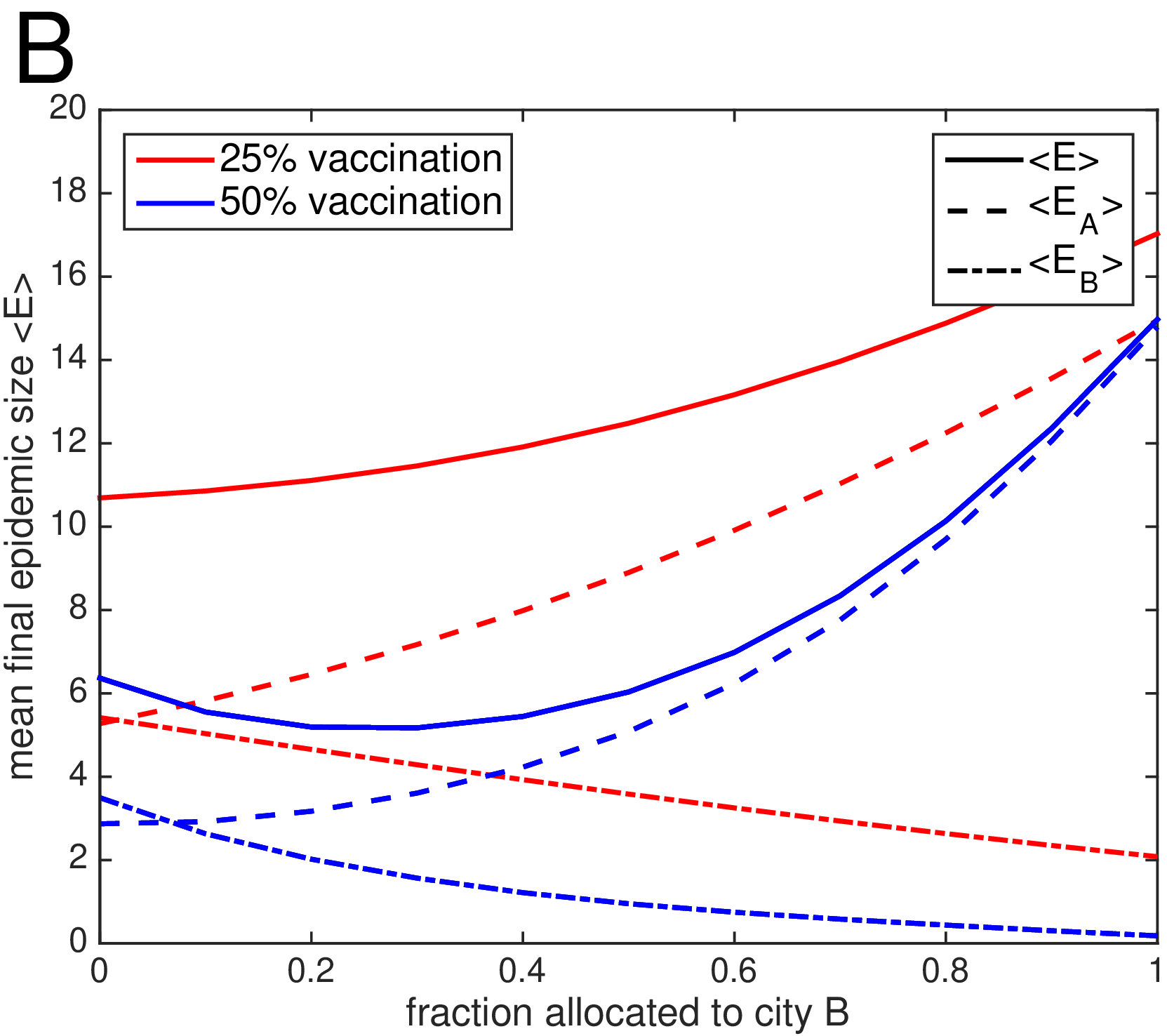}
\label{specex}}
\caption{{\bf Mean final epidemic size as a function of the fraction of vaccine allocated to city B.} The stochastic combined mean final epidemic size $\left<E\right>$ as a function of the fractional allocation to city B is plotted in (a) for different fixed amounts of available vaccine. The specific cases of 25\% and 50\% vaccination are highlighted in (b), where mean final epidemic sizes for individual populations $\left<E_A\right>$ and $\left<E_B\right>$ are also plotted, in addition to the combined final size $\left<E\right> = \left<E_A\right> + \left<E_B\right>$. The time delay is set to $\tau = 5$ days and the coupling $\fab = 0.05$. City A has 39 initial susceptibles and 1 infective; city B has 40 initial susceptibles. The recovery rate $\gamma = 0.15$ and the reproductive number $r_0 = 2$.}
\label{fig:example}
\end{figure}

To more closely examine the stochastic results, the mean final epidemic sizes $\left<E\right>$ are plotted in Fig.\@ \ref{ex} as a function of the fraction of vaccine allocated to city B in the specific case of $\tau = 5$ and $\fab = 0.05$. Here, the difference in mean epidemic size between the worst-case and optimal scenarios is relatively large, since time delay and coupling are both low. It is evident that larger amounts of vaccine generally result in smaller mean epidemic sizes. When there is enough vaccine for 12.5\% of the population, the minimum of the line occurs when no vaccines are given to city B. The minimum shifts rightward as the amount of vaccine increases; when there is enough vaccine for 87.5\% of the population, the minimum occurs for a 0.5 fractional allocation. Note that the maxima always occur for all vaccine allocated to city B.

It is possible for a smaller amount of vaccine to more efficiently reduce final epidemic size when allocated strategically than a larger amount of vaccine that is improperly used. If the worst-case protocol is followed when there are 40 or greater vaccines (50\% or more), all vaccines would be allocated to city B, and at most 40 individuals, or the entire population of city B, can be vaccinated. Any remaining vaccine would be wasted. This would result in a mean final epidemic size $\left<E\right> \approx 15$. However, when the optimal protocol is followed for 20 vaccines (25\%), all 20 vaccines are given to city A, and $\left<E\right> \approx 11$, less than the previously described case. Even though fewer individuals are vaccinated, fewer individuals become infected. Targeting the source of the outbreak in order to minimize the chance of infection spreading elsewhere is a better strategy than preemptively vaccinating those who do not live nearby. 

The specific cases of 25\% and 50\% vaccination are highlighted in Fig.\@ \ref{specex}, where the individual mean final epidemic sizes $\left<E_A\right>$ and $\left<E_B\right>$ are plotted alongside the combined mean final epidemic size. The epidemics are predictably smaller in city B. For both cases, $\left<E_A\right>$ monotonically increases as a function of the vaccine allocated to city B, while $\left<E_B\right>$ monotonically decreases as a function of vaccine allocation. $\left<E\right>$ is also monotonic for 25\% vaccination; however, $\left<E\right>$ for 50\% vaccination is non-monotonic, with a minimum that lies at around 0.3 fractional allocation. 

For the parameter space explored so far in this paper, $\left<E_A\right>$ is minimized for vaccine allocations that give all or most vaccine to city A, and $\left<E_B\right>$ is minimized for vaccine allocations that give all or most vaccine to city B. That is, it is never beneficial for an individual population to donate the majority of its vaccines to another population, even if the outbreak is occurring elsewhere. This is also true for the deterministic model. However, for both populations viewed as a whole, vaccine donating or vaccine sharing is often desirable.

\begin{figure}[h!]
\includegraphics[width=0.75\textwidth]{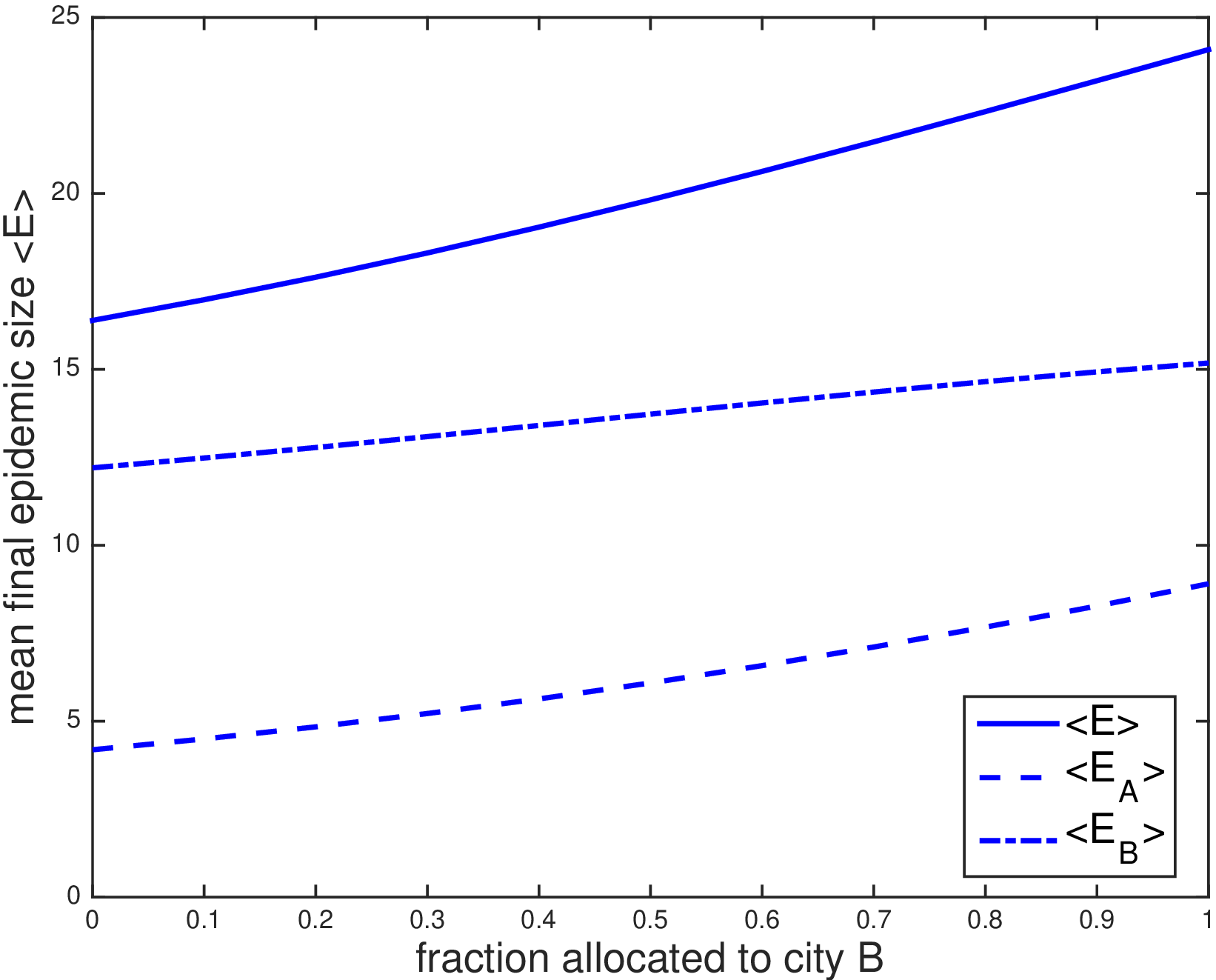}
\caption{{\bf A case in which both cities benefit the most when all vaccine is allocated to city A.} The combined mean final epidemic size $\left<E\right>$ and the individual mean final sizes $\left<E_A\right>$ and $\left<E_B\right>$ are minimized when all vaccines are allocated to city A in order to eradicate the infection at its source, due to the low coupling ($\fab = 0.01$) and low time delay ($\tau = 5$). 10 vaccines are given to city A, which has a population of 20 individuals, while city B, which has a population of 100, remains completely susceptible.}
\label{fig:diffsize}
\end{figure}

For unequal population sizes, it is possible for both city A and city B to individually benefit when all vaccine is given to city A, provided that city A is relatively small compared to city B and time delay and coupling are very low. In this case, the probability of infection spreading from A to B is low, such that a relatively small amount of vaccine will be effective at mitigating the infection in city A. Thus, it is reasonably safe for B to remain vulnerable and instead focus efforts on eradicating the epidemic in city A. This case is illustrated in Fig.\@ \ref{fig:diffsize}, where there are now 20 individuals (one of which is infected) in city A and 100 susceptibles in city B. The individual mean final epidemic sizes $\left<E_A\right>$ and $\left<E_B\right>$ are both minimized when all 10 available vaccines are allocated to city A.

\section*{Discussion}
Using an SIR model to simulate epidemics in interacting populations, we analyzed the tradeoffs involving vaccination, time delay, and coupling in order to determine optimal resource allocation strategies. The earlier a population undergoes mass vaccination, the more effective the intervention, and less vaccine will be required to eradicate the epidemic. For interacting cities, the optimal solution might favor one city over another, depending on the time delay until vaccine is deployed. In general, weaker coupling leads to protocols which allocate vaccine in more disparate proportions to the two cities, while stronger coupling favors protocols which allocate vaccine in equal (and proportional) amounts to each city. 

\subsection*{Synchrony} 
\begin{figure}[h!]
\subfloat{\includegraphics[width=0.49\textwidth]{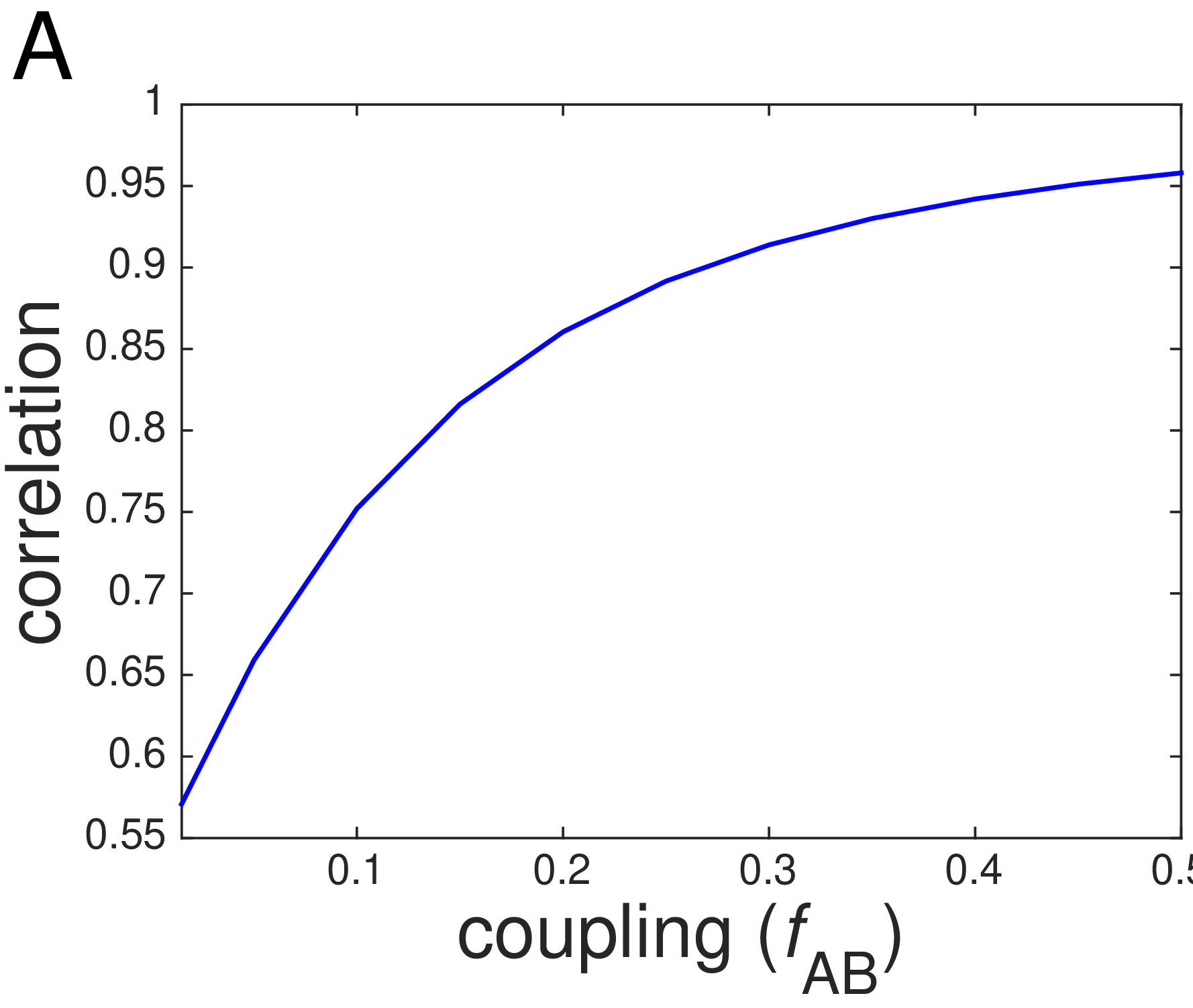}
\label{corr}}\ \ \
\subfloat{\includegraphics[width=0.49\textwidth]{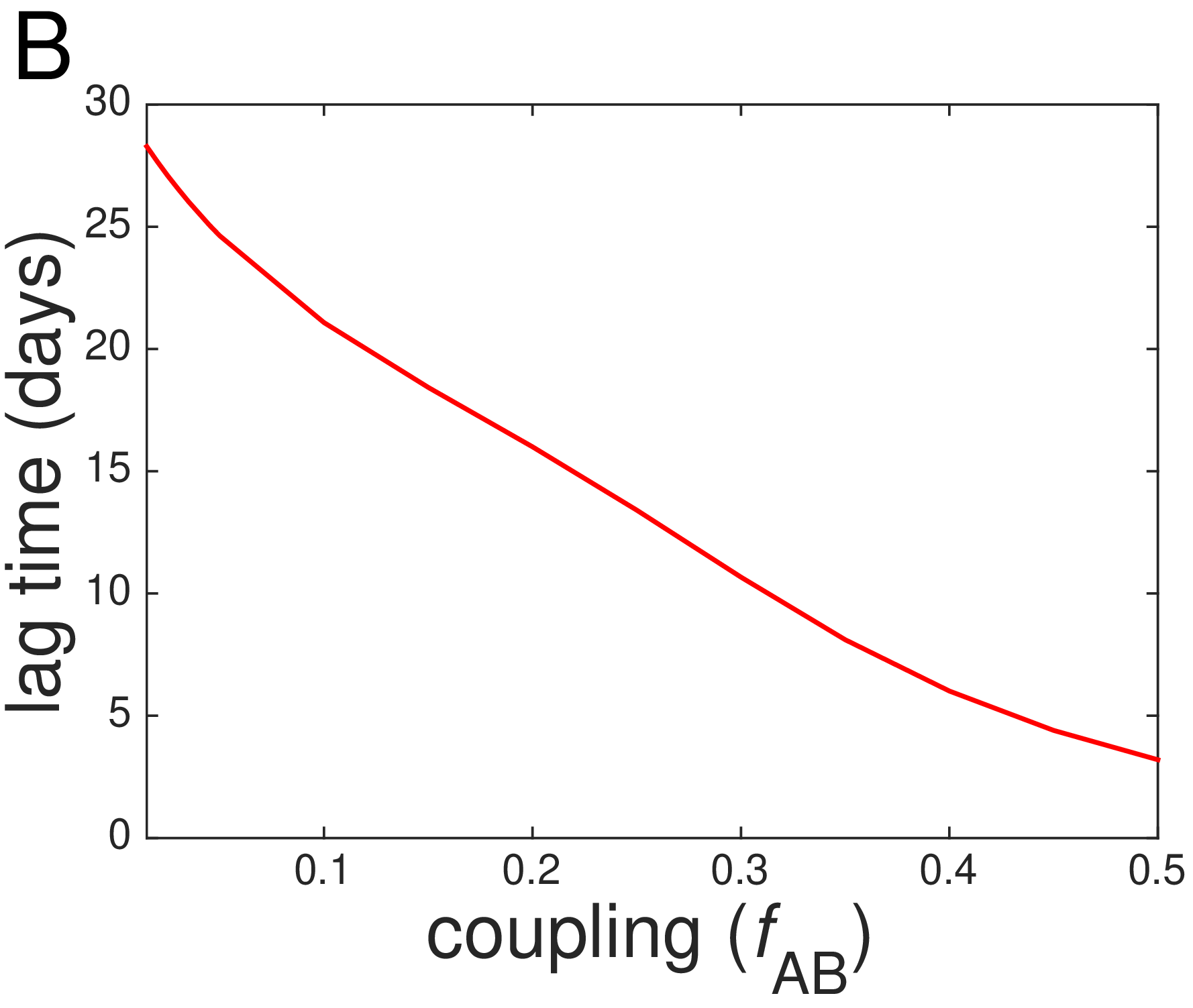}
\label{lag}}
\caption{{\bf Synchrony of epidemics represented by correlation and lag time as function of coupling.} A: The correlation between the number of infectives $I(t)$ in each city is plotted as a function of coupling $\fab$. Higher correlation corresponds to higher synchrony. B: The lag time between epidemics is plotted as a function of $\fab$. A lower lag time corresponds to higher synchrony. For both plots, 25\% of the total combined population is vaccinated after 5 days in an equal allocation, such that each city receives 10 vaccines. City A has 39 initial susceptibles and 1 infective; city B has 40 initial susceptibles. The recovery rate $\gamma = 0.15$ and the reproductive number $r_0 = 2$.}
\label{fig:sync}
\end{figure}

A useful quantity to examine is synchrony, a measure of the correlation between the infection dynamics of interacting populations, which increases as a function of coupling. This relationship between correlation and coupling is demonstrated for the specific case of $\tau = 5$ and 25\% vaccination in \ref{corr}. Here, correlation is defined as the Pearson correlation coefficient between the number of infectives in each city, $I_A(t)$ and $I_B(t)$, as measured over a full simulation. Synchrony also affects the optimal vaccine allocation; higher synchrony tends to result in a more equal division of vaccine between the two cities, while lower synchrony tends to result in a less equal allocation.

The effects of vaccination on synchrony can depend on the specific parameters of the disease. 
In general, vaccination increases the probability of disease extinction while also reducing the effective coupling, since transmission is decreased. Reduced coupling leads to reduced synchrony, which in turn reduces the probability of permanent widespread extinction: asynchrony across spatial scales increases the probability that infection will return to a population where the disease was previously eradicated, in what is termed a ``rescue event" \cite{kr}. It has been observed (for example, during the measles outbreak in England and Wales in the 1970s and 19780s) that the effects of reduced infection due to vaccination approximately cancel those of asynchrony \cite{kr}. 

However, Rohani et al. \cite{synchrony} comprehensively compared measles and pertussis outbreaks in England and Wales between 1940 and 1990, finding that while vaccination decreased synchrony of measles outbreaks, it actually increased synchrony of pertussis outbreaks. This was attributed to the relatively longer pertussis infection period, as well as the increased age at vaccination for pertussis, which implies increased individual movement and therefore increased coupling.

Pulsed vaccination, in which the population is vaccinated periodically over a specified timeframe, in conjunction with steady mass vaccination, has been suggested as a method to synchronize epidemics, thereby more effectively controlling the disease \cite{earn}. Pulsed vaccination is particularly useful for vaccinating children with an immunization course administered over a certain age range. 

Large time delay and low coupling contribute to asynchrony, as illustrated in the first panel of Fig.\@ \ref{fig:opt2}. When $\fab \lesssim 0.25$ and $\tau \gtrsim 30$, the optimal allocation highly favors city B. At this late point in the epidemic, the infection has begun to die out in city A but has started to grow in city B, due to a \textit{lag time} between the epidemics in each city. We define the mean lag time as the mean difference in time between when the number of infectives $I_A(t)$ in city A reaches a maximum and when $I_B(t)$ in city B reaches a maximum. The mean lag time decreases as a function of coupling, as illustrated in Fig.\@ \ref{lag}, since higher coupling results in more synchronous epidemics. If there is a localized outbreak in city A, the increased lag time at low coupling could be relatively advantageous to city B by effectively buying time for vaccine to be transported and administered to city A.

\subsection*{Broader Implications} While the communities simulated in this paper are relatively small, we have qualitatively illustrated tradeoffs that we expect to be important at larger scales, including at the level of entire nations. 
For example, if resources have been stockpiled, it can be undesirable to keep all vaccines within a nation, especially if frequent international travel encourages the spread of infection. On the other hand, if coupling is extremely low, vaccinating an uninfected population while an outbreak occurs elsewhere can be disadvantageous. Rather, exporting vaccine to the location of the outbreak in order to localize the spread of the infection could more effectively reduce its severity.

Even for large populations, there can be a significant probability that the disease does not propagate between cities or countries. Should the amount of available vaccine be limited, it can be disadvantageous, from a holistic perspective, for an uninfected population to vaccinate itself when an epidemic is occurring in another population. As described in Fig.\@ \ref{fig:example}, vaccinating the uninfected population preemptively can lead to more severe epidemics. Vaccine is typically allocated in proportion to regional population sizes, as with the H1N1 vaccine within the United States during the 2009 epidemic \cite{cdcflu}. When the disease is widespread, this is likely the optimal solution, but if the infection is relatively localized, it might be favorable to focus on eradicating the disease near its source.

The model described in this paper is based on the basic SIR framework, but it can readily be made more complex by adding compartments, such as an $E$ class representing exposed individuals who are symptomatic but not infectious. Other variants on the SIR model include, for example, the SIRS model, in which a recovered individual stays immune for some finite amount of time before reverting to susceptibility, or the SIS model, which skips the recovery stage altogether. The choice of model will be dictated by the characteristics of the specific disease being studied. 

The methods detailed in this paper can also be applied to a more realistic model incorporating, for instance, different rates of infection based on age. Spatial separation and heterogeneity could be explicitly included, as well as specific logistic considerations such as delays due to transport of vaccine from storage facilities to clinics. Other forms of epidemic control can be included, such as transportation restrictions or quarantining. Restricting transportation between cities will reduce the effective coupling $f$ and reduce the overall movement of individuals.
Quarantining serves to reduce the effective infection period $1/\gamma$ by isolating infectives from the rest of the population \cite{kr}. Quarantine effectively dampens the infectious spread and is an attractive method of epidemic control, since it does not involve the high costs of producing and distributing vaccine. However, unlike vaccination, quarantine does not directly immunize susceptibles, but indirectly protects them by lowering the probability of transmission.

Moreover, while we do not consider them in this paper, SIR models on networks are also commonly used to model epidemics and are especially useful for understanding spatial dynamics. An example SIR network model could have cities defined on nodes with edge weights derived from transportation frequency, such as in the model devised by Matrajt et al \cite{birdflu}. In this model, the spread of infection within cities is described with the deterministic SIR equations, while interactions between cities is described by a stochastic process. While this hybrid model includes the probability of non-spreading, the deterministic intra-city interactions do not account for the variability of the disease dynamics and the probability of early disease extinction within a community.

The problem of optimizing dynamic resource allocation can be also be applied to natural disasters, oil spills, and other circumstances which are of interest to policy officials. SIR-type models similar to those formulated in this paper can in particular be applied to wildfires, with compartments of \textit{unburned}, \textit{burning}, and \textit{burned} corresponding to susceptible, infected, and recovered, respectively \cite{nada}. While the concept of coupled populations in epidemiology does not have a direct analogy in the realm of natural disasters, wildfires may break out in rapid succession in different areas, requiring quick and optimized allocation of resources.

The difference between the outcomes of optimal and worst-case scenarios can be compared to identify where intervention is most effective. Furthermore, it may not be preferred to follow the optimal protocol in certain cases. For example, if the optimal solution prescribes vaccinating only city A, leaving city B's populace fully vulnerable to the infection is likely unfavorable to the residents of city B, especially since the probability of the epidemic spreading to city B is nonzero. In such a case, policymakers could take the worst-optimal difference into consideration to choose a vaccination protocol such that a middle ground is struck between a mathematically optimal solution and a realistic, but not excessively deleterious, solution.

\section*{Acknowledgments}
The authors would like to acknowledge Edwin Yuan for helpful discussions. This work was supported by the David and Lucile Packard Foundation and the Institute for Collaborative Biotechnologies through contract no. W911NF-09-D-0001 from the U.S. Army Research Office.

\bibliography{refs}
\bibliographystyle{unsrt}

\clearpage
 
\setcounter{figure}{0}
\renewcommand{\thefigure}{S\arabic{figure}}

\section*{Supporting Information}
\subsection*{Algorithmic considerations}

In this paper, we used a stochastic SIR framework to model the spread of an epidemic between two interacting cities, performing numerical integration with the implicit-Euler (IE) method \cite{jg}. An alternative method, Krylov subspace approximation (KSA), can also be applied to the master equation; KSA is commonly used to simplify Markov chains containing exponentials of large, sparse matrices \cite{sidje}. The KSA method uses a small, dense matrix to efficiently approximate solutions to the master equation, rather than directly integrating with the generator matrix itself. The master equation can be written as the recursive series of equations:
\be \vec{P}(t_i) = \exp(\tau \mathds{A}) \vec{P}(t_{i-1}), \ee
which can then be solved using, for instance, the Expokit package in MATLAB \cite{expokit}.
The approximate solutions take the form:
\be \vec{P}(t_i) = ||\vec{P}(t_{i-1})||_2\ \mathds{V}(t_{i-1})\ \exp\{\tau \mathds{H}(t_{i-1})\} \ \hat{e}_1, \ee
where the columns of $\mathds{V}$ form an orthonormal basis of the $m$-dimensional Krylov subspace:
\be \mathcal{K}_m(t) = \text{span}\{\vec{P}(t), \tau \mathds{A} \vec{P}(t), \hdots, (\tau \mathds{A})^{m-1} \vec{P}(t)\}. \ee
$\mathds{H}$ is a dense $m \times m$ upper-Hessenberg matrix, generated using the Arnoldi method, which satisfies:
\be \mathds{V}^{\text{T}} \tau \mathds{A} \mathds{V} = \tau \mathds{H}, \ee
and $\hat{e}_1$ is the first column of the $m$-dimensional identity matrix. Because $m << K$, where $K$ is the dimension of $\mathds{A}$, the upper-Hessenberg matrix $\mathds{H}$ is much smaller than $\mathds{A}$. 

The accuracy of the IE method is controlled by the size of the time step, which is chosen by the user; the error of the IE method is $\mathcal{O}(\tau)$. On the other hand, the accuracy of the KSA method is controlled by the size of the Krylov subspace and the Expokit error tolerance, both of which are specified by the user. Furthermore, the KSA method is prone to instabilities arising from compounded errors, and occasionally cannot produce a normalized, nonnegative probability vector without resorting to a heuristic approximation. In general, the KSA method is less computationally intensive than the IE method with the chosen time step size, but also less accurate.

A 200-day simulation of two cities with 40 people each, 0.25 coupling, and 40 vaccines administered at a 10 day delay, executed in MATLAB 8.5 on a 2.5 GHz Intel Core i7 processor, took 874 seconds with the IE method using a time step of 0.01 seconds. In comparison, the same simulation took 344 seconds with the KSA method using a Krylov subspace of dimension 65 and an Expokit error tolerance of $1 \times 10^{-3}$. An ``exact" solution was generated by running the KSA method with a subspace of dimension 85 and an error tolerance of $1 \times 10^{-7}$, which took 440 seconds. The results generated with the IE and KSA methods were compared to the exact solution and had L2 errors of order $10^{-4}$ and $10^{-2}$, respectively. 
While the time required to run the ``exact" KSA simulation is about half that required for the IE method, the IE method allows for greater control of the integration process. The time steps are smaller and uniform in size and allow for more temporal dynamics to be extracted.

\begin{figure}[h!] \centering
\includegraphics[width=0.5\textwidth]{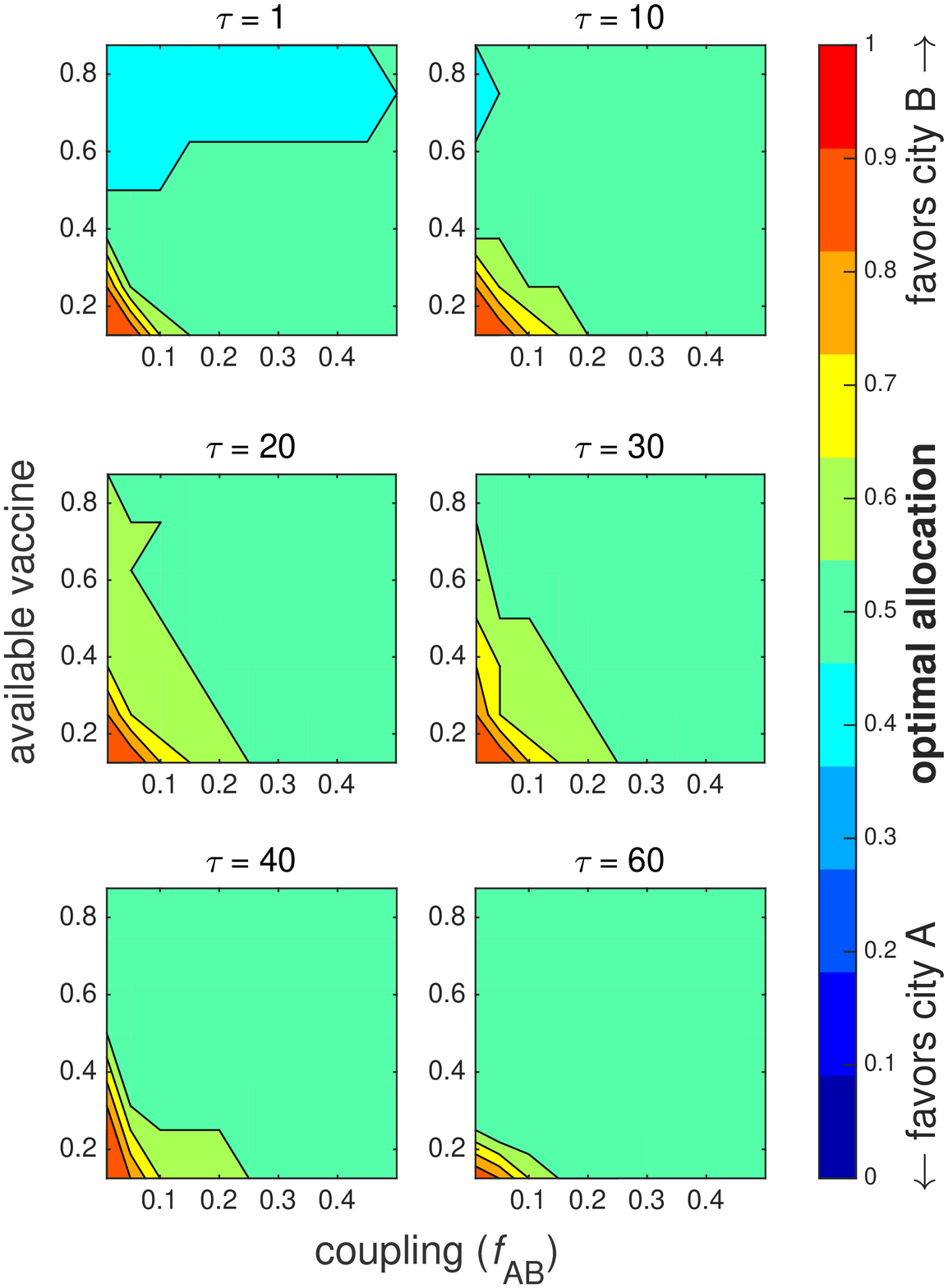}
\label{fig:S1_Fig}
\caption{{\bf Deterministic optimal strategies at different time delays.} The optimal fraction of total vaccine allocated to city B in the deterministic model is plotted as a function of available vaccine (expressed as a fraction of the total combined population) and coupling $f_{\text{AB}}$ for different fixed values of time delay $\tau $ranging from 1 to 60 days.}
\end{figure}

\begin{figure}[h!] \centering
\includegraphics[width=0.5\textwidth]{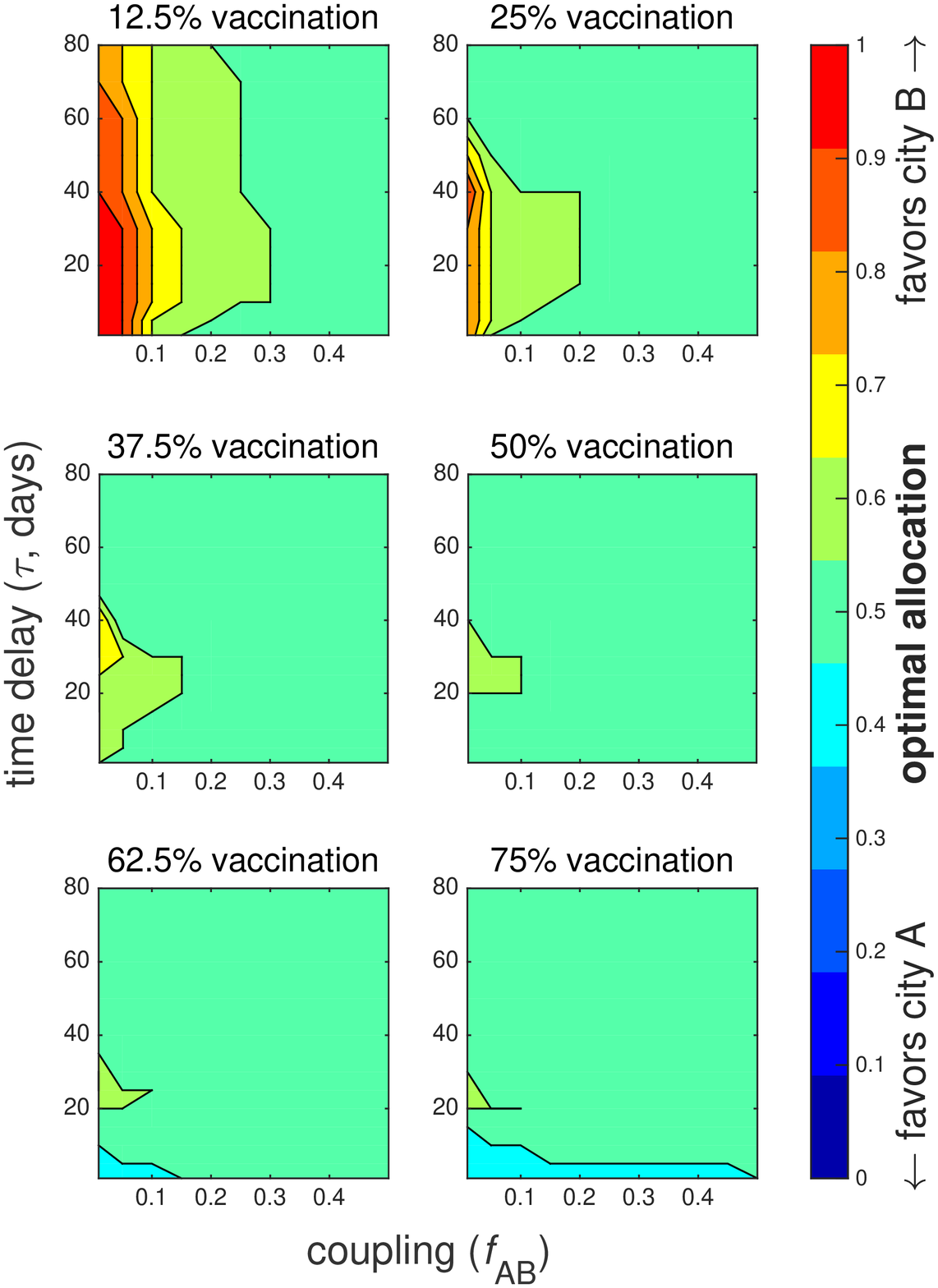}
\label{fig:S2_Fig}
\caption{{\bf Deterministic optimal strategies for different amounts of available vaccine.} Optimal fraction of total vaccine allocated to city B in the deterministic model plotted as a function of time delay $\tau$ and coupling $f_{\text{AB}}$ for different fixed values of available resources ranging from 10 to 60 vaccines (i.e.\@ 12\% to 75\% vaccination).}
\end{figure}

\begin{figure}[h!] \centering
\includegraphics[width=0.5\textwidth]{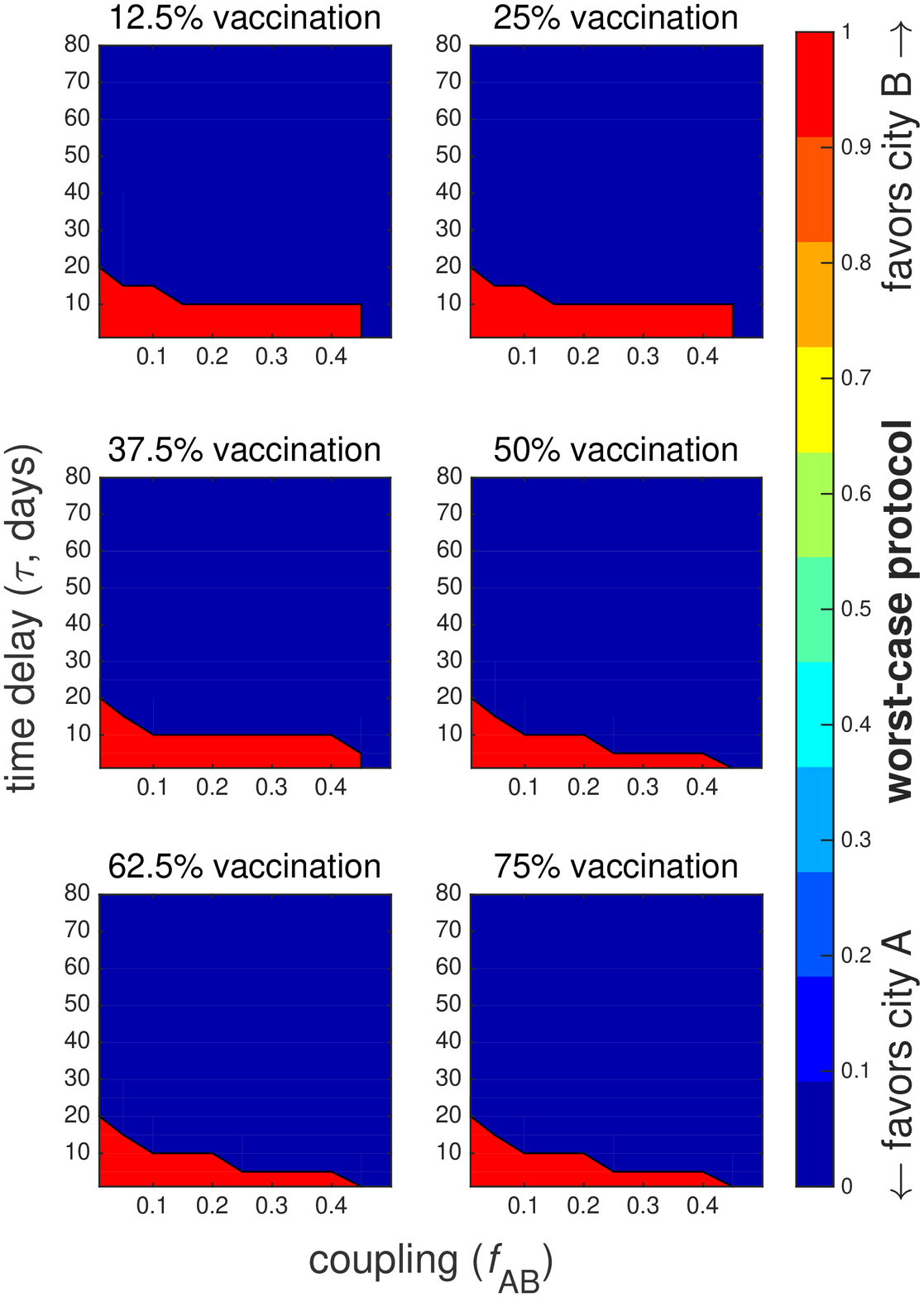}
\label{fig:S3_Fig}
\caption{{\bf Worst-case vaccine allocations in the stochastic model.} The vaccine allocations resulting in maximum mean final epidemic size $\left<E\right>^{\text{max}}$ are plotted as a function of time delay $\tau$ and coupling $f_{\text{AB}}$ for different fixed amounts of available vaccine. City A has 39 initial susceptibles and one infective; city B has 40 initial susceptibles. The recovery rate $\gamma = 0.15$ and the reproductive number $r_0 = 2$.}
\end{figure}

\begin{figure}[h!] \centering
\includegraphics[width=0.5\textwidth]{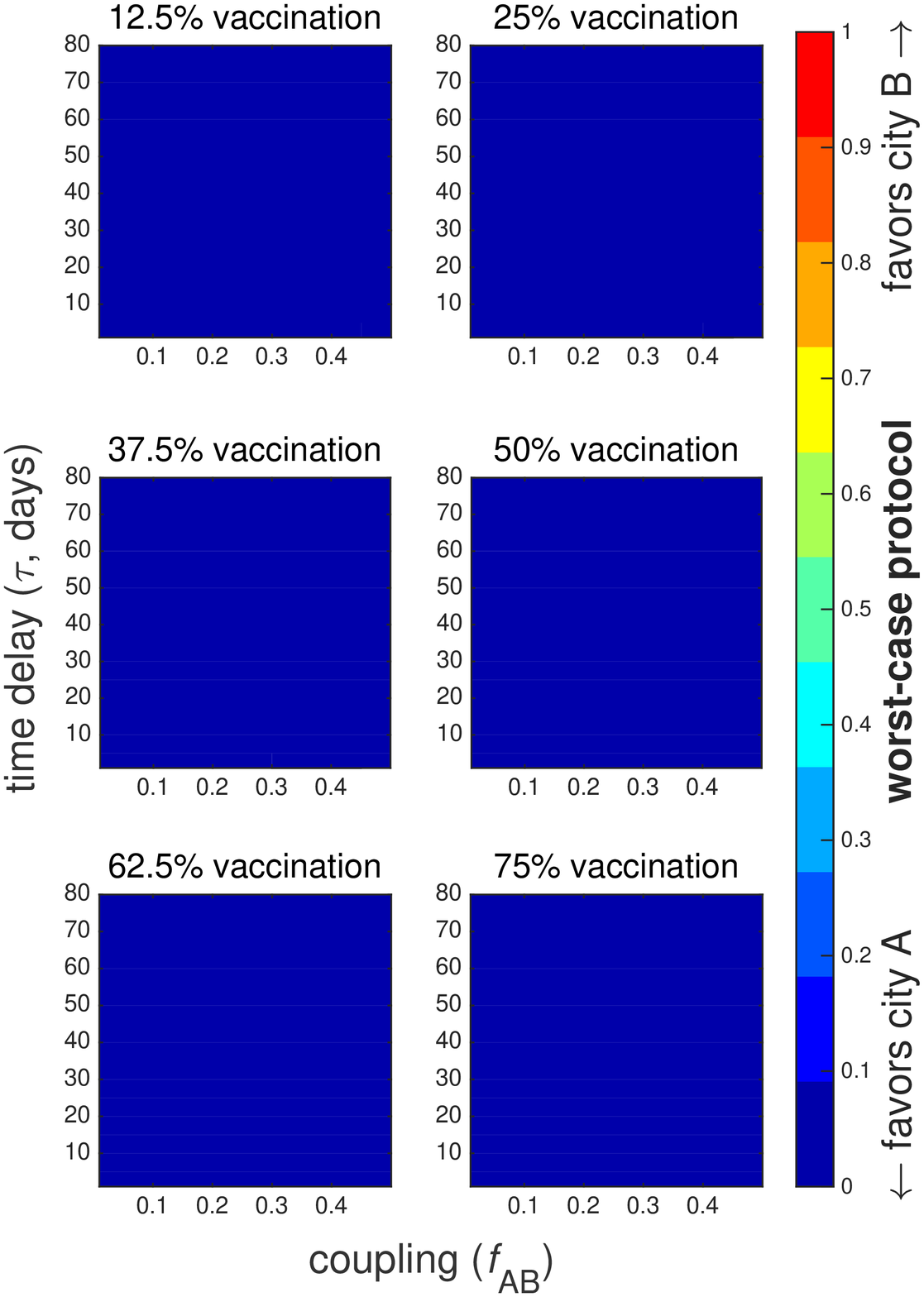}
\label{fig:S4_Fig}
\caption{{\bf Worst-case vaccine allocations in the deterministic model.} The vaccine allocations resulting in maximum final epidemic size $E_{\text{det}}^{\text{max}}$ are plotted as a function of time delay $\tau$ and coupling $f_{\text{AB}}$ for different fixed amounts of available vaccine. City A has 39 initial susceptibles and one infective; city B has 40 initial susceptibles. The recovery rate $\gamma = 0.15$ and the reproductive number $r_0 = 2$.}
\end{figure}

\begin{figure}[h!] \centering
\includegraphics[width=0.5\textwidth]{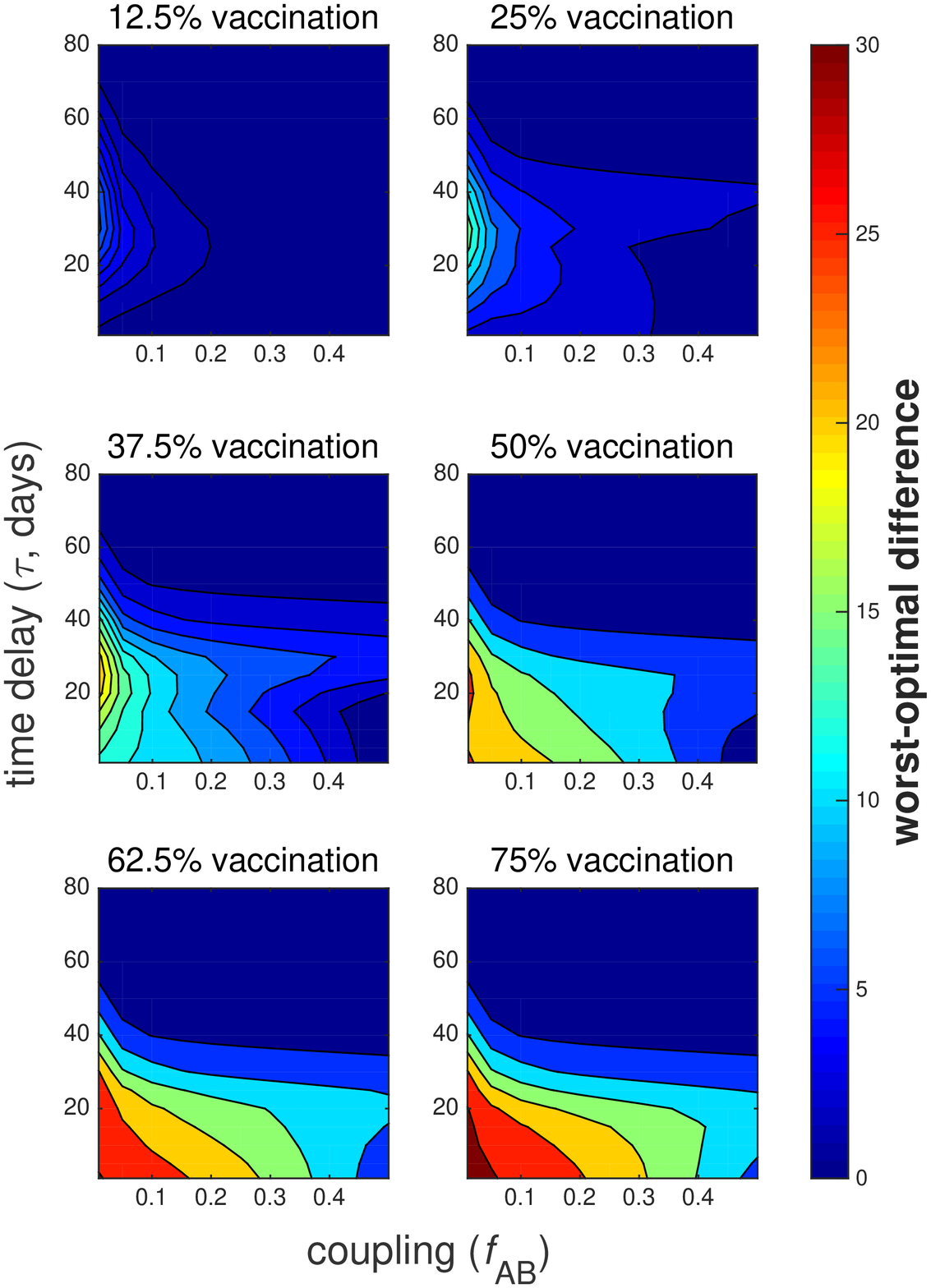}
\label{fig:S5_Fig}
\caption{{\bf Difference between deterministic optimal and worst-case final epidemic sizes.} The difference in deterministic final epidemic size $E_{\text{det}}$ between worst-case and optimal protocols is plotted as a function of time delay $\tau$ and coupling $f_{\text{AB}}$ for different fixed amounts of available vaccine.}
\end{figure}

\end{document}